\documentclass[journal]{IEEEtran}
\usepackage{indentfirst}
\usepackage{multirow}
\usepackage{graphicx}
\usepackage{color}
\usepackage{xcolor}
\usepackage{colortbl}
\usepackage{listings}
\usepackage{enumerate}
\usepackage{subfigure}
\usepackage{amsmath,amsthm,amscd,amssymb,latexsym,upref,stmaryrd}
\usepackage[noend]{algpseudocode}
\usepackage{algorithmicx,algorithm}
\usepackage{multirow}
\usepackage{xcolor}
\usepackage{amsfonts,epsfig}
\usepackage{CJK}
\usepackage{float}
\usepackage{cite}
\usepackage{subfigure}
\usepackage{lipsum}
\usepackage{multicol}
\usepackage{setspace}
\usepackage{bm}
\usepackage{stfloats}
\usepackage{booktabs}
\usepackage{threeparttable}

\def\BibTeX{{\rm B\kern-.05em{\sc i\kern-.025em b}\kern-.08em
		T\kern-.1667em\lower.7ex\hbox{E}\kern-.125emX}}

\begin{document}
		\title{Angle-based SLAM on 5G mmWave Systems: Design, Implementation, and Measurement}
	\author{Jie~Yang, Chao-Kai~Wen, Jing~Xu, Hang~Que, Haikun~Wei, and Shi~Jin
	\thanks{Jie~Yang is with the School of Automation and the Frontiers Science Center for Mobile Information Communication and Security Southeast University, Nanjing, China (e-mail: yangjie@seu.edu.cn).
	Shi~Jin, Jing~Xu, and Hang~Que, are with the National Mobile Communications Research Laboratory, Southeast University, Nanjing, China (e-mail: \{jinshi;xujing\_;quehang\}@seu.edu.cn). Chao-Kai~Wen is with the Institute of Communications Engineering, National Sun Yat-sen University, Kaohsiung, 804, Taiwan (e-mail: chaokai.wen@mail.nsysu.edu.tw).
	Haikun Wei is with the Key Laboratory of Measurement
	and Control of CSE, Ministry of Education, School of Automation,
	Southeast University, Nanjing, China (e-mail:hkwei@seu.edu.cn).}}
	\maketitle	
	
\begin{abstract}
Simultaneous localization and mapping (SLAM) is a key technology that provides user equipment (UE) tracking and environment mapping services, enabling the deep integration of sensing and communication. The millimeter-wave (mmWave) communication, with its larger bandwidths and antenna arrays, inherently facilitates more accurate delay and angle measurements than sub-6 GHz communication, thereby providing opportunities for SLAM. However, none of the existing works have realized the SLAM function under the 5G New Radio (NR) standard due to specification and hardware constraints. In this study, we investigate how 5G mmWave communication systems can achieve situational awareness without changing the transceiver architecture and 5G NR standard. We implement 28 GHz mmWave transceivers that deploy OFDM-based 5G NR waveform with 160 MHz channel bandwidth, and we realize beam management following the 5G NR. Furthermore, we develop an efficient successive cancellation-based angle extraction approach to obtain angles of arrival and departure from the reference signal received power measurements. On the basis of angle measurements, we propose an angle-only SLAM algorithm to track UE and map features in the radio environment. Thorough experiments and ray tracing-based computer simulations verify that the proposed angle-based SLAM can achieve sub-meter level localization and mapping accuracy with a single base station and without the requirement of strict time synchronization. Our experiments also reveal many propagation properties critical to the success of SLAM in 5G mmWave communication systems.
\end{abstract}

	
\section{Introduction} \label{sec:introduction}

Integrated sensing and communication (ISAC) is emerging as a key feature of 6G wireless communications, enabling the simultaneous realization of massive device connection, high-speed data transmission, and high-precision sensing \cite{6G0,6G1}. The exploitation of high frequencies, wide bandwidths, and the dense deployment of massive antenna arrays further enable the implementation of ISAC techniques. Extensive theoretical studies have demonstrated the integration and coordination gains of ISAC. In particular, the use cases of ISAC can be summarized into four categories \cite{ISAC}:
(1) high-accuracy localization and tracking;
(2) simultaneous imaging, mapping, and localization;
(3) augmented human sense; and
(4) gesture and activity recognition.
The ubiquitous Internet of Things architecture for ISAC was studied in \cite{ISAC3}. Signal processing techniques and fundamental limits for ISAC were reviewed in \cite{ISAC-SP} and \cite{I2}, respectively. A comprehensive overview of the background, range of key applications, and state-of-the-art approaches of ISAC was provided in \cite{ISAC4}.

Experimental demonstrations and prototypes for ISAC are currently urgent and vital to transition ISAC from theory to practice. Radio-based user-centric indoor mapping has demonstrated the excellent prospects of mobile environment mapping in future millimeter-wave (mmWave) networks \cite{rs1}. The system operates at the 28 GHz band, deploying OFDM-based 5G New Radio (NR) uplink waveform with 400 MHz channel bandwidth.
In \cite{rs2}, a fully digital wideband joint communication-radar (JCR) system was developed to detect objects by extending an mmWave communication setup with an additional full-duplex radar receiver and capturing the single-input multiple-output JCR channel with a moving antenna on a sliding rail. Prototypes of high-resolution mmWave and THz imaging on portable devices were realized in \cite{ISAC} and \cite{THz}, respectively, using virtual array scanning mechanisms.
In addition to localization, tracking, and imaging, hand gesture recognition was investigated by WiFi devices \cite{ZDQ} and 60 GHz mmWave ISAC prototype systems in \cite{rs3}. The results justify the possibility of high-resolution sensing while considering the physical aperture constraint of a typical mobile device.

Most previous ISAC prototypes have been limited to implementing radar functions. In this study, we consider the potential of simultaneous localization and mapping (SLAM) in mmWave communication systems. SLAM has relatively mature applications in the field of robotics, and it is often achieved by leveraging the robot's sensors, such as an inertial measurement unit (IMU), cameras, or lasers \cite{RSS,vslam,lidar}. These sensors provide more landmarks than the anchors available in typical radio networks.
The concept of radio-based SLAM is to use the multipath channel information to locate radio nodes, reconstruct images of the propagation environment, and provide reliable and accurate user localization service simultaneously. A computationally efficient radio-based SLAM method is proposed in \cite{radioSLAM}, which is promising for realizing real-time UE localization and environment mapping. The development of ISAC prototype systems that can perform radio-based SLAM is a timely and emerging area of research.
Experimental results in \cite{rs4} demonstrated the excellent performance of SLAM in challenging indoor environments by using synthetic measurements and real ultra-wideband radio signals. The challenges of SLAM are attributed to diffuse multipath propagation, unknown feature association, and limited visibility of features. However, time of arrival (TOA) measurements require strict network synchronization techniques, and achieving synchronization accuracy required for sensing may be difficult in cellular networks \cite{RTT}. Recent studies have developed SLAM algorithms using angle measurements based on the same principles as \cite{rs4}.
Angle of arrival (AOA), angle of departure (AOD), and TOA measurements are used in \cite{mmwavesw}. Moreover, AOA, TOA, and the component signal-to-noise ratio (SNR) of the multipaths are utilized in \cite{eslam1,eslam2,eslam3}, and the probability of detection is connected to the component SNR of the multipaths for the first time \cite{eslam1,eslam2}. The work in \cite{rs6} validated radio-based SLAM via 60 GHz WLAN, which simulates sector-level beam sweeping with the rotation of the horn antenna. However, the implementation of SLAM in real mmWave systems by reusing the 5G NR beam management process has not been reported yet, which is the main focus of the current study.

The highly directional transmission characteristic of mmWave is inherently promising for SLAM services \cite{r17,r19,r20}. On the one hand, mmWave links are highly susceptible to rapid channel variations \cite{mmwave}. The current release of 3GPP 5G NR adopts a beam management framework based on beam sweeping, measurements, and reporting \cite{3GPPBM,b0}. Thus, angle information, such as AOA and AOD, can be collected during beam management processes inherent to mmWave networks at no additional overhead of positioning reference signals. Moreover, unlike TOA measurements, angle measurements do not require strict time synchronization, which is difficult to achieve in communication systems. Recent studies have demonstrated effective beam selection and AOA estimation methods in mmWave networks, which are built on the uplink sounding reference signal transmissions by employing the latest 5G NR specifications \cite{channelest}.
On the other hand, multipath angle parameters are strongly correlated with environmental geometry due to the sparsity of mmWave frequencies and the presence of specular reflections \cite{mmwavesw,mms,mms2}. Thus, the reflective path can be regarded as a signal from the virtual base station (BS), as shown in Fig. \ref{fig:floorplan}, which is the mirror image of the physical BS on the reflective surface. The physical and virtual BSs are static despite the movement of the reflection point with the UE. Therefore, we regard physical and virtual BSs as physical anchor (PA) and virtual anchor (VA) \cite{rs4,mmwavesw,CSLAM2,yj5,jy}, respectively. PA and VA are considered radio features because they act as landmarks and describe the radio environment. Moreover, PA and VA simplify the description of the radio environment and enable feature association in SLAM algorithms. Thus, realizing SLAM, which localizes user equipment (UE) and maps PA/VA in the propagation environment, is promising through the 5G NR beam management process.

However, achieving SLAM under the existing 5G NR beam management framework faces many challenges. First, the time-frequency resources are limited. The beam sweeping process is confined to 5 ms for one frame according to \cite{3GPPBM}. Thus, the number of beam pairs measured during a half frame is limited. Second, mmWave hardware is constrained. The beam angular resolution of practical hardware-constrained mmWave systems is restricted due to the limited number of antenna elements and phase shifter bits.
\textbf{Commercial mmWave 5G networks have already been launched, but the potential for sensing that can be achieved by hardware and resource-constrained 5G mmWave communication systems remains unresolved.}
To address these challenges, we develop a proof-of-concept mmWave platform for angle-based SLAM.
Instead of changing the traditional transmitters of 5G NR communication systems, we integrate radio sensing into current communication-only cellular networks by multiplexing the beam sweeping process \cite{r9,r10,r11,r13,r14}. Our approach achieves sub-meter level localization and mapping accuracy despite the time-frequency resource limitations and hardware constraints. Our main contributions are presented as follows:
\begin{itemize}
	\item
	\textbf{Platform implementation:}
    We implement a $28$ GHz mmWave ISAC platform under the existing 5G NR beam management framework. The platform includes the mmWave phased array, mmWave local oscillator (LO), software-defined radio (SDR), reference clock node, and reference clock source. The SDR supports a bandwidth of $160$ MHz. Each mmWave phased array covers approximately $120^{\circ}$ with eight discrete beam directions. The grid error of the adopted beam codebook is $6^{\circ}$ to $9^{\circ}$. We realize the beam management process according to the 3GPP 5G NR frame structure and obtain reference signal received power (RSRP) measurements.

	\item
	\textbf{Algorithm design:}	
    We propose an angle-based SLAM algorithm to track UE and map features in the radio environment. The algorithm comprises three techniques: Angle-Extract, Angle-SLAM, and IMU-Calib. First, we extract the AOAs and AODs from the real RSRP measurements using a successive cancellation-based angle extraction method. Second, we extend the classic belief propagation (BP) SLAM algorithm \cite{rs4} to achieve SLAM with only angle measurements. Finally, we use the IMU to calibrate the estimation of the UE track because IMU provides accurate short-time measurements. The proposed algorithm fully multiplexes the beam management process of communications without consuming additional communication resources.
		
	\item
	\textbf{Real measurement:}
    We conduct comprehensive experiments with the proposed hardware and software designs. The results demonstrate that the proposed angle-based SLAM algorithm can achieve sub-meter level localization and mapping accuracy in most experimental scenarios. In complex scenarios with nonsmooth scatterers, with beam birth and death, and without the assistance of IMU, the proposed algorithm can still achieve meter-level localization and mapping capabilities.  Multiuser collaboration can improve localization and mapping performance by approximately 50\% and 77\%, respectively, compared to cases without collaboration. IMU can help improve the mapping performance of extreme VAs by approximately 50\% and 30\% for simulation and experiment, respectively, compared to cases without IMU assistance. Additionally, our measurements reveal many important propagation properties critical to the success of SLAM.
\end{itemize}

The rest of this paper is organized as follows.
Section \uppercase\expandafter{\romannumeral2} introduces the system model and problem formulation.
We describe the experimental hardware platform in Section \uppercase\expandafter{\romannumeral3}.
The experimental software platform is explained in Section \uppercase\expandafter{\romannumeral4}.
Our experimental results are presented in Section \uppercase\expandafter{\romannumeral5}, and we conclude the study in Section \uppercase\expandafter{\romannumeral6}.

\section{System Model}\label{s2}

We consider a 5G NR-compatible mmWave MIMO communication system with a static BS and a moving UE, as illustrated in Fig. \ref{fig:floorplan}.
BS and UE use analog beamforming and time-division multiplexing beam sweeping, respectively, to maintain highly directional transmission/reception and collect radio features of the propagation environment simultaneously.

\subsection{Geometric Model}\label{gm}
\begin{figure}
	\centering
	\includegraphics[scale=0.8]{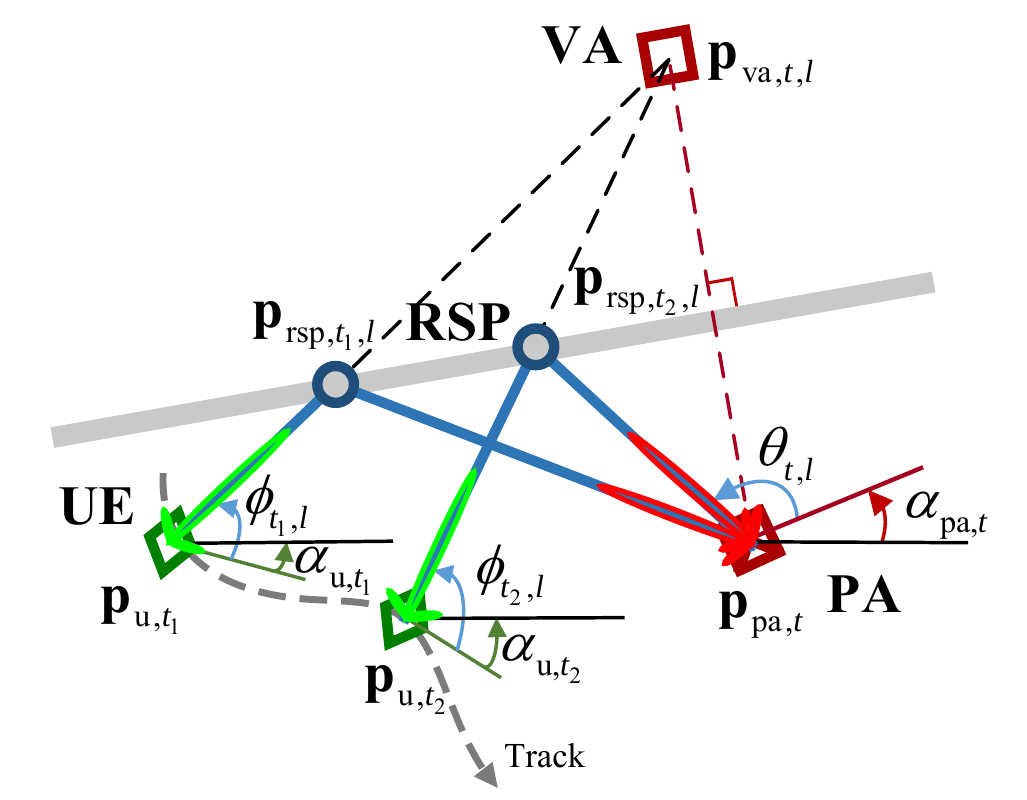}
	\caption{Geometric relationship among UE, PA, VA, and RSP, denoted by green diamond, red crossed box, red square, and blue circle, respectively. The track of the UE is depicted by a gray dotted line. The reflective surface is presented by a gray solid line.} \label{fig:floorplan}\vspace{-0.2cm}
\end{figure}

The state of the UE at time $t$ is denoted as ${\mathbf{u}_{t} = [\mathbf{p}_{{\rm u},t},\mathbf{v}_{{\rm u},t}]}$, where ${\mathbf{p}_{{\rm u},t}=[x_{{\rm u},t},y_{{\rm u},t}]}$ indicates the location, and ${\mathbf{v}_{{\rm u},t}=[\dot{x}_{{\rm u},t},\dot{y}_{{\rm u},t}]}$ represents the velocity. The BS is considered as a static anchor and hence is referred to as PA.
The true location of PA at time $t$ is represented by ${\mathbf{p}_{{\rm pa},t}=[x_{{\rm pa},t},y_{{\rm pa},t}]}$. We estimate the location of PA at time $t$, hence the subscript $t$.
The true location of the \textbf{p}oint on the \textbf{r}eflective \textbf{s}urface (RSP) is denoted as ${\mathbf{p}_{{\rm rsp},t,l}=[x_{{\rm rsp},t,l},y_{{\rm rsp},t,l}]}$, where ${l=2,\ldots,L_{t}}$ and ${L_{t}-1}$ represents the number of reflective surfaces. The location of the reflection point on the same reflective surface changes as the UE moves, as shown in Fig. \ref{fig:floorplan}.  For a single-bounce specular NLOS path, the virtual source of the corresponding path is represented by VA, which is the mirror image of PA on the reflective surface. The true location of the $l$-th VA at time $t$ is denoted as $\mathbf{p}_{{\rm va},t,l}$, where ${l=2,\ldots,L_{t}}$. The single-bounce specular NLOS paths on the same reflective surface originate from the same VA. Thus, VA can be viewed as an abstract feature that describes the reflective surface. Therefore, the radio features of the propagation environment include the locations of PAs and VAs.

We consider the downlink (DL) beam measurements and denote the $l$-th AOD and AOA as $\theta_{t,l}$ and $\phi_{t,l}$ at time $t$, respectively.
Let $\alpha_{{\rm u},t}$ and $\alpha_{{\rm pa},t}$ represent the orientation of the UE and PA, respectively. \footnote{We consider a 2D configuration to simplify the algorithm realization and experimental process. Our main goal is to investigate how 5G mmWave communication systems can achieve situational awareness without changing the transceiver architecture and 5G NR standard. As we focus on indoor experimental scenarios, a 2D configuration can easily be satisfied by setting the BS and UE to the same height. If the UE and BS are at different altitudes and can obtain 3D AOAs and AODs, the proposed algorithm can still work by using only the azimuth of AOAs and AODs. In our future work, we will consider extending the study to 3D scenarios, where UPA will be deployed.}
We obtain the geometric relationship of UE, PA, and RSP as
\begin{equation}\label{1}
\tan(\tilde{\theta}_{t,l}) = \dfrac{y_{{\rm rsp},t,l}-y_{{\rm pa},t}}{x_{{\rm rsp},t,l}-x_{{\rm pa},t}},
\end{equation}
\begin{equation}\label{2}
\tan(\tilde{\phi}_{t,l}) = \dfrac{y_{{\rm rsp},t,l}-y_{{\rm u},t}}{x_{{\rm rsp},t,l}-x_{{\rm u},t}},
\end{equation}
where ${\tilde{\theta}_{t,l}=\theta_{t,l}+\alpha_{{\rm pa},t}}$ and ${\tilde{\phi}_{t,l}=\phi_{t,l}+\alpha_{{\rm u},l}}$.
Afterward, the geometric relationship of PA, RSP, and VA is presented as
\begin{equation}\label{3}
\mathbf{p}_{{\rm va},t,l} = \mathbf{p}_{{\rm rsp},t,l} + \| \mathbf{p}_{{\rm rsp},t,l} - \mathbf{p}_{{\rm pa},t} \|\bm{\phi}_{t,l},
\end{equation}
where $\bm{ \phi}_{t,l} = [\cos(\tilde{\phi}_{t,l}), \sin(\tilde{\phi}_{t,l})]$ and $\| \cdot \|$ denotes the 2-norm.

The geometric model in \eqref{1}-\eqref{3} suggests that the initial locations of the UE and PA are required to determine the VAs, RSPs, and subsequent UE locations when only angle measurements are used.
To estimate the RSPs, we can use the AOA and AOD measurements at ${t=1}$ using \eqref{1} and \eqref{2}, respectively. We can then convert the RSPs into VAs using \eqref{3}. The PAs and VAs act as landmarks of the radio environment, which can aid in the subsequent data fusion and localization of the moving UE.


\begin{figure*}
	\centering
	\vspace{-0.2cm}
	\includegraphics[scale=0.60]{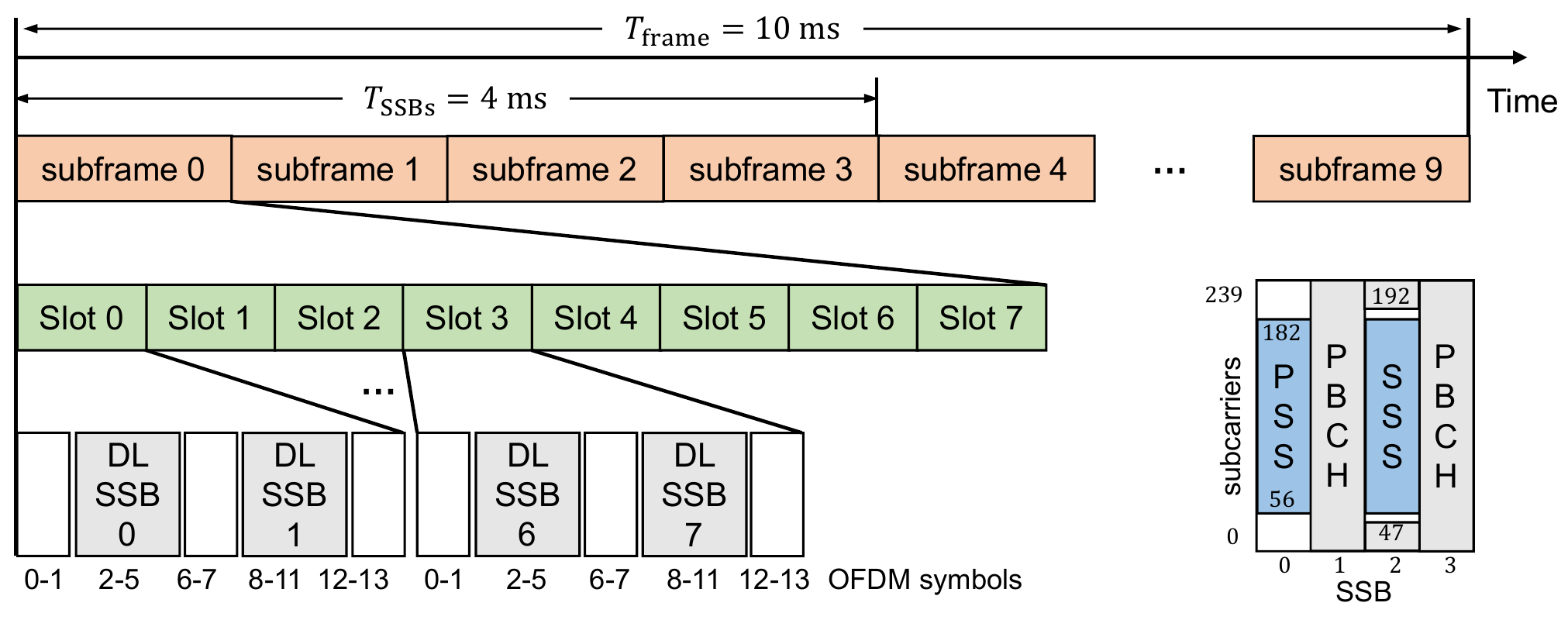}
	\caption{Frame structure and SSB composition. A frame lasts ${T_{\rm frame}=10}$ ms, with $10$ subframes at $1$ ms each.
	A subframe comprises of $8$ slots, and a slot comprises $14$ OFDM symbols. } \label{fig:SSB}
	\vspace{-0.2cm}
\end{figure*}
\begin{figure*}
	\centering
	\vspace{-0.1cm}
	\hspace{-0.5cm}
	\includegraphics[scale=0.60]{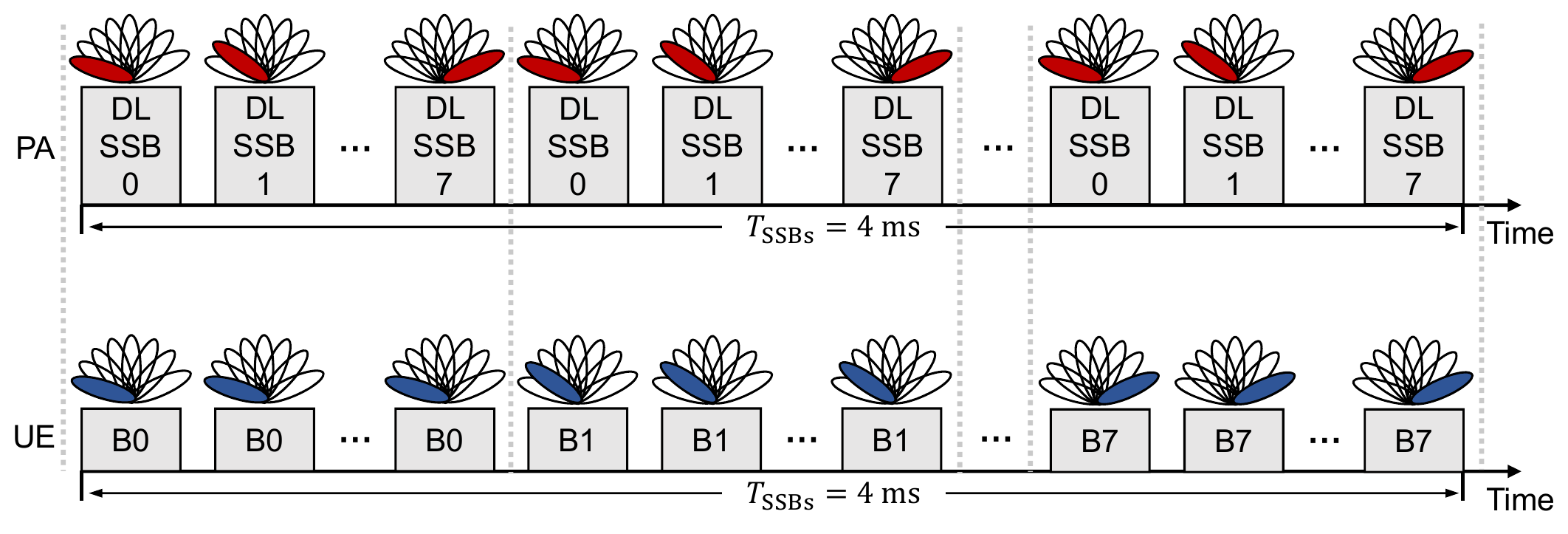}
	\caption{Procedure of exhaustive beam sweeping, where eight beams are considered on UE and PA sides.} \label{fig:ES}
	\vspace{-0.4cm}
\end{figure*}

\subsection{Signal Model}\label{BM}
The DL measurements for beam management in this study follow the 3GPP specifications Release 15 \cite{3GPPBM}. The OFDM waveform subcarrier spacing is given by ${15\times 2^\gamma}$ kHz, where ${\gamma\in \mathbb{Z}}$ and ${\gamma\leqslant 4}$. A frame has a duration of ${T_{\rm frame}=10}$ ms and consists of $10$ subframes of $1$ ms each. A slot consists of $14$ OFDM symbols, and the number of slots in a subframe depends on the numerology used, given that the symbol duration is inversely proportional to the subcarrier spacing. An example of the frame structure used in this study is shown in Fig. \ref{fig:SSB}. A synchronization signal block (SSB) consists of four OFDM symbols in time and $240$ subcarriers in frequency, and comprises a primary synchronization signal (PSS), a secondary synchronization signal (SSS), and a physical broadcast channel (PBCH). The PBCH can be used to estimate the reference signal received power (RSRP) of the SSB. DL beam management includes four different operations.

\subsubsection{Beam Sweeping}\label{sweeping}
Each SSB can be assigned to a specific beam direction, with multiple SSBs grouped into a burst that covers a spatial area with predetermined intervals and directions.
The maximum duration of an SSB burst is limited to ${T_{\rm SSBs}\leqslant 5}$ ms per frame. An example of the exhaustive beam sweeping method is shown in Fig. \ref{fig:ES}. The SSBs are transmitted using $B_{\rm TX}$ beam directions on the PA side in a polling manner on $N_{\rm s}$ active subcarriers. The synchronization signal transmitted at the $i$-th beam direction is denoted as ${\mathbf{x}_{n} = {s}_{n}\mathbf{a}_{\rm TX}(\theta_i)}$, where ${s}_n$ denote the synchronization signal at the $n$-th subcarrier in the OFDM symbol, ${i=0,\ldots,B_{\rm TX}-1}$, and $\mathbf{a}_{\rm TX}(\cdot)$ is the steering vector. A set of DL SSBs is periodically transmitted by the PA, with different periodicities ${\Delta T = \{5,10,20,40,80,160\}}$ ms.

\subsubsection{Beam Measurement}
During this process, the UE evaluates the quality of the received signal. To perform exhaustive beam sweeping, beam 0 is first fixed for receiving on the UE side in order to measure the RSRPs of $B_{\rm TX}$ different beams sent by the PA. Subsequently, beam 1 to ${B_{\rm RX}-1}$ on the UE side is fixed sequentially for receiving. The signal transmitted by the PA at the $i$-th beam and received by the UE at the $j$-th beam is given by
\begin{equation}
{y}_{n,i,j} = \mathbf{a}^{\rm H}_{\rm RX}(\phi_j)\mathbf{H}_{n}\mathbf{a}_{\rm TX}(\theta_i){s}_{n} + {\varrho},
\end{equation}
where ${ j=0,\ldots,B_{\rm RX}-1}$, $\mathbf{a}_{\rm RX}(\cdot)$ is the steering vector, $(\cdot)^{\text{H}}$ represents the conjugate transpose of a vector/matrix, $\mathbf{H}_{n}$ is the channel between UE and PA at the $n$-th subcarrier, and $\varrho$ is the additive Gaussian noise.
Therefore, the $(i,j)$-th RSRP is defined as\footnote{As the noise correlation of adjacent sub-carriers is small, we multiply adjacent sub-carriers to avoid noise effects.}
\begin{equation}
r_{i, j} = \dfrac{1}{N_s-1}\left\lvert \sum\limits_{n=0}^{N_s-2} \dfrac{{y}_{n,i,j} {y}_{n+1,i,j}^{*} }{s_ns_{n+1}^{*}} \right\rvert,
\end{equation}
where $(\cdot)^{*}$ and $\lvert \cdot \rvert$ represent the conjugate and modulus of a complex value, respectively.

\begin{figure*}
	\centering
	\vspace{-0.4cm}
	\includegraphics[scale=0.4]{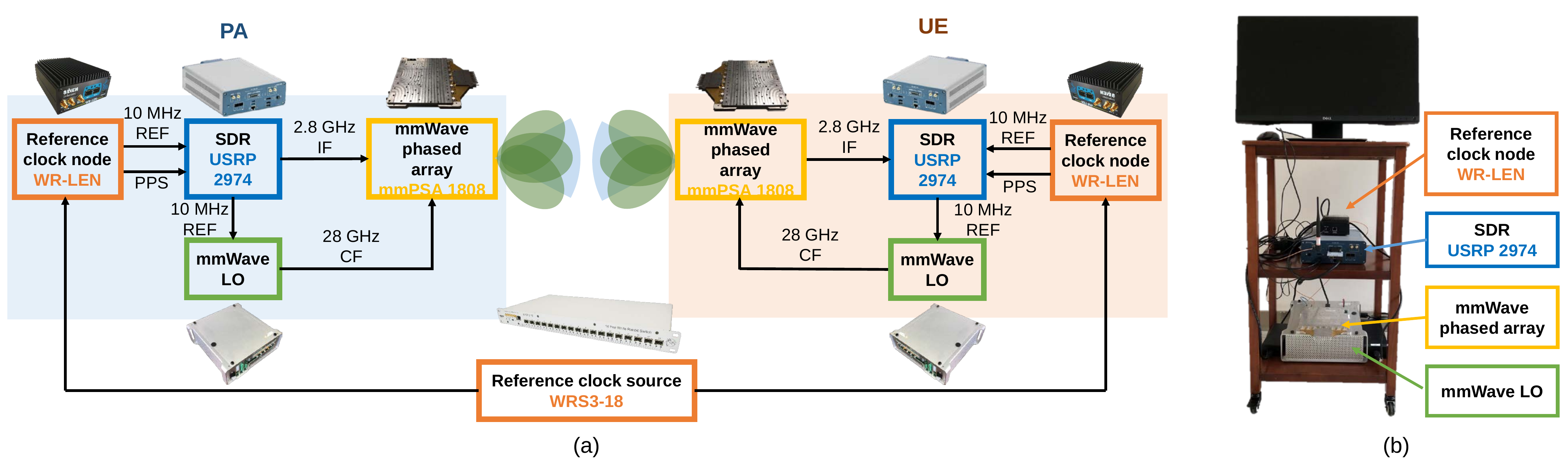}
	\caption{(a) Hardware connections of the mmWave ISAC platform. (b) Picture of one UE or PA.} \label{fig:hardware}
	\vspace{-0.1cm}
\end{figure*}

\subsubsection{Beam Determination}
The UE selects suitable beam or beams for communication based on the measurements obtained with the beam measurement procedure. In this study, the UE compares RSRPs to find the optimal transmitted and received beam pairs. In contrast to traditional single-functional systems, the effective beam pairs of multipaths can also realize the sensing function of the propagation environment in a dual-functional ISAC system, as the multipath component information contains the location and state of radio features in the environment.
In this study, the UE obtains angle information of the determined beam pairs, and the estimated AOD and AOA of the $l$-th path are denoted as $\hat{\theta}_{t,l}$ and $\hat{\phi}_{t,l}$, respectively, for $l=1,\ldots,\hat{L}_{t}$.
We can define the stacked angle measurement vector as:
\begin{equation}
\mathbf{z}_{t}=\left[\left(\hat{\theta}_{t,1},\hat{\phi}_{t,1}),\cdots,(\hat{\theta}_{t,\hat{L}_{t}},\hat{\phi}_{t,\hat{L}_{t}}\right)\right].
\end{equation}
The UE can obtain a sequence of measurements $\mathbf{z}_{1:T}=[\mathbf{z}_{1},\ldots,\mathbf{z}_{T}]$ through accumulation of $T$ times.

\subsubsection{Beam Reporting}
The UE transmits beam quality and decision information to the PA during the initial access process in the corresponding time slots.

\subsection{Problem Formulation}
This study aims to enable angle-based SLAM by reusing beam measurements in a 5G NR-compatible mmWave MIMO system. The localization and environment mapping are carried out on the UE side. The scenario under consideration involves a situation where the UE enters an unfamiliar indoor environment, and its tracks are unknown due to GPS signal blockage indoors. The goal of SLAM is to determine the positions of PAs, VAs, RSPs, and UE tracks based on angle measurements $\mathbf{z}_{1:T}$. We assume that the initial positions of UEs and PAs are known, given that only angle information is available. The UE gradually establishes the environment geometry relative to its start point (initial position).

\section{Experimental Hardware Platform}\label{HP}	

We implement an mmWave ISAC platform in this study, and the commodity devices used are described in this section.

\begin{figure}
	\vspace{-0.4cm}
	\centering
	\includegraphics[scale=0.54]{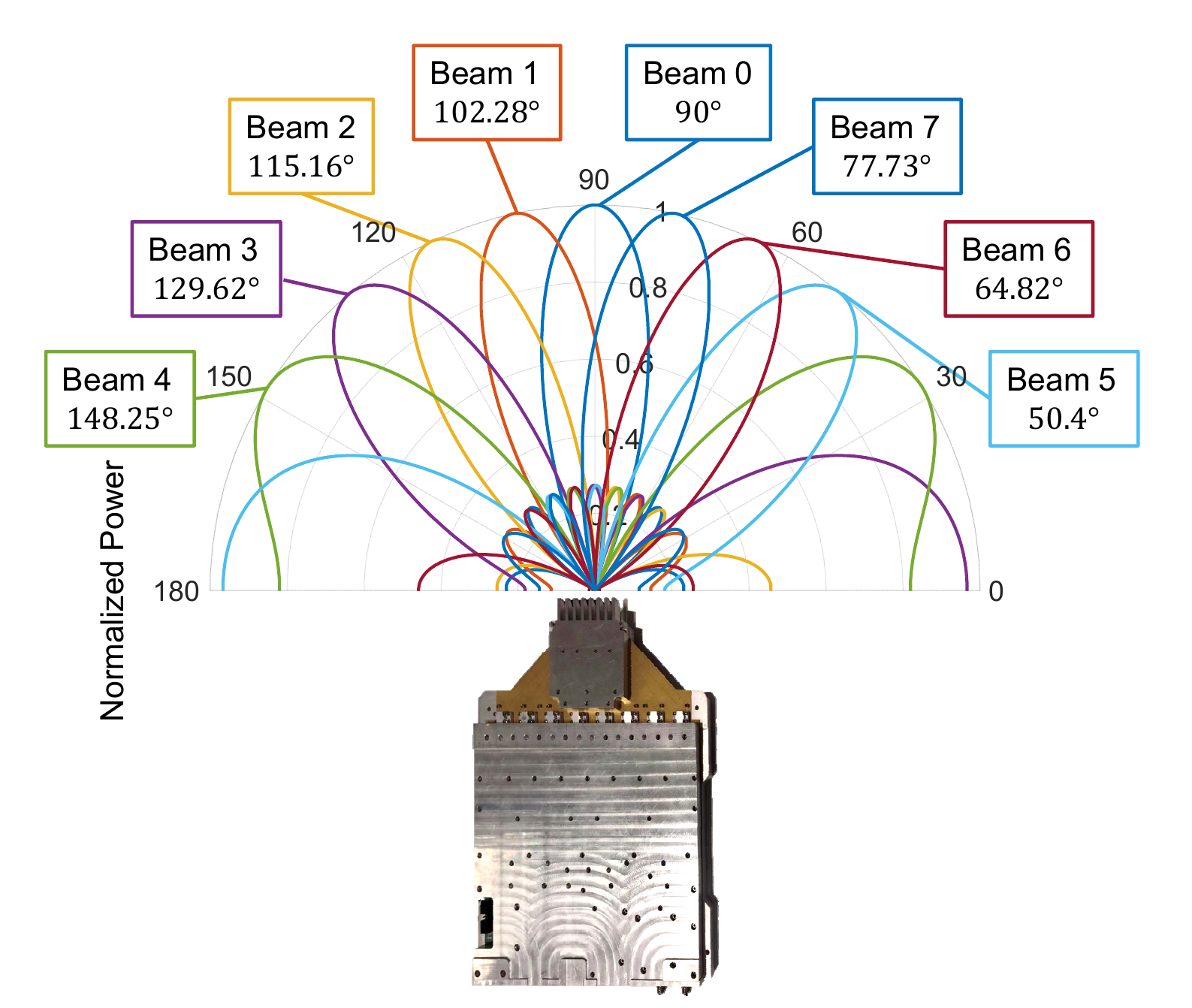}
	\caption{Theoretical beam pattern of the designed codebook, in which eight beams cover from $50.4^{\circ}$ to $148.25^{\circ}$.} \label{fig:codebook}
	\vspace{-0.4cm}
\end{figure}

\subsection{Hardware Composition}
Our mmWave ISAC platform consists of various components, including the mmWave phased array (mmPSA-1808), the SDR (USRP-RIO 2974), the mmWave LO, the reference clock node (WR LEN), and the reference clock source (WRS3-18). Fig. \ref{fig:hardware} illustrates the architecture of the mmWave ISAC platform. The reference clock source and node provide a $10$ MHz reference clock signal for PA and UE, respectively. The SDR supports a bandwidth of $160$ MHz and allows for intermediate frequency (IF) up/down conversion. The mmWave LO module is connected to the mmWave TX/RX head for up/down conversion to a $28$ GHz carrier frequency (CF). The mmWave TX/RX head is then connected to the mmWave phased array for over-the-air transmission.

For DL measurements of beam management, the PA first uses the SDR to generate the baseband signal, which is then upconverted to a $2.8$ GHz IF signal. The IF signal is upconverted to a $28$ GHz CF signal by the mmWave LO and transmitted by the mmWave phased array with analog precoding. The receiving process on the UE side is the opposite of the transmitting process on the PA side. The control signal of the mmWave phased array is generated by the GPIO port of the USRP-RIO 2974 to realize beam switching in strict accordance with the corresponding 5G NR SSB resource (as illustrated in Fig. \ref{fig:SSB}).

\subsection{mmWave Phased Array and Beam Codebook}
The mmWave phased array is a linear array consisting of eight antenna elements and one radio frequency chain. The working frequency range is ${27-29}$ GHz, with a center frequency of $28$ GHz, and a TDD transceiver duplex mode. The antenna spacing is $6.3$ mm (${d=0.588\lambda}$), and each antenna element is connected with a $6$-bit phase shifter, allowing for the formation of ${2^6=64}$ different phases. The phase interval is ${\Delta \psi = 360^{\circ} /64=5.625^{\circ}}$, and the beam switching time is less than $300$ ns.

Let $N_{\rm TX}$ and $N_{\rm RX}$ denote the number of active antenna elements.
The $\beta$-th vector of the DFT beam codebook is $[1, e^{-j\beta\Delta \psi},\ldots,e^{-j(N_{\rm *}-1)\beta\Delta \psi}]$, where ${\beta=0,\ldots,63}$, and ${\rm *}$ denotes ${\rm TX}$ or ${\rm RX}$.
In this study, we have chosen ${B_{\rm TX}=B_{\rm RX}=8}$ beams with a phase difference of $ {\psi_b = 8b\Delta \psi}$, where ${b=0,\ldots,B_{\rm *}-1}$.
Therefore, the corresponding ${8\times 8}$ beam codebook matrix is given by
\begin{equation*}
\begin{bmatrix}
1&1&1& \ldots & 1\\
1&e^{j8\Delta \psi}&e^{j16\Delta \psi}& \ldots & e^{j56\Delta \psi}\\
\vdots&\vdots&\vdots& \ddots & \vdots\\
1&e^{j8(\!N_{\rm *}\!-\!1\!)\Delta \psi}&e^{j16(\!N_{\rm *}\!-\!1\!)\Delta \psi}& \ldots & e^{j56(\!N_{\rm *}\!-\!1\!)\Delta \psi}
\end{bmatrix}.
\end{equation*}
The normalized beam pattern is shown in Fig. \ref{fig:codebook}, where ${N_{\rm TX}=N_{\rm RX}=4}$. \footnote{As the beam interval is larger than $12^{\circ}$ (Fig. \ref{fig:codebook}), we form wide beams with four active antenna elements to cover the target space.}
We observe that eight beams cover approximately $120^{\circ}$, and the angle space between adjacent beams gradually increases as the beams move away from the normal direction of the antenna array.
Therefore, the beam direction (in rad) according to Fig. \ref{fig:codebook} is given as
\begin{equation}\label{book} 
\Theta \!=\! [\underbrace{50.4^{\circ}}_{b=5}, \underbrace{64.8^{\circ}}_{6}, \underbrace{77.7^{\circ}}_{7},\underbrace{90^{\circ}}_{0}, \underbrace{102.3^{\circ}}_{1},  \underbrace{115.2^{\circ}}_{2}, \underbrace{129.6^{\circ}}_{3}, \underbrace{148.3^{\circ}}_{4}],
\end{equation}
which is defined as the angle dictionary. The elements in $\Theta$ are sorted by the angle from small to large. The array orientation should be compensated continuously when using the angle dictionary $\Theta$.

\subsection{Frame Structure}
The frame structure used in this study is shown in Fig. \ref{fig:SSB}, following the 5G NR standard. One frame is $10$ ms and consists of $10$ subframes. Each subframe comprises eight slots, with each slot containing $14$ OFDM symbols. The FFT length is set to $1024$, and the cyclic prefix (CP) length of the OFDM symbol $0$ at slots $0$ and $4$ is $136$, while that of the remaining OFDM symbols is $72$. The subcarrier spacing is $120$ kHz, and the sampling rate is $122.88$ MSps, with the number of effective subcarriers for data transmission set to $792$.

One SSB consists of four OFDM symbols, as shown in Fig. \ref{fig:SSB}. The first symbol is PSS, which is the m-sequence with a length of $127$, and the third symbol is SSS, which is the gold-sequence with a length of $127$. For subframes $0$ to $3$, the system inserts two SSBs in each slot, starting from OFDM symbols $2$ and $8$, while the remaining OFDM symbols are available for communication. Subframe $4$ is reserved for the physical random access channel (PRACH), with the UE using the optimal uplink transmit beam for beam reporting. All OFDM symbols can be used for communication for subframes $5$ to $9$. Therefore, there are $64$ SSBs in one $10$ ms frame, and the system takes only $4$ ms to perform a complete ${8\times 8}$ beam sweeping.

\section{Experimental Software Platform}\label{SP}	

\begin{figure}
	\centering
	\includegraphics[scale=0.45]{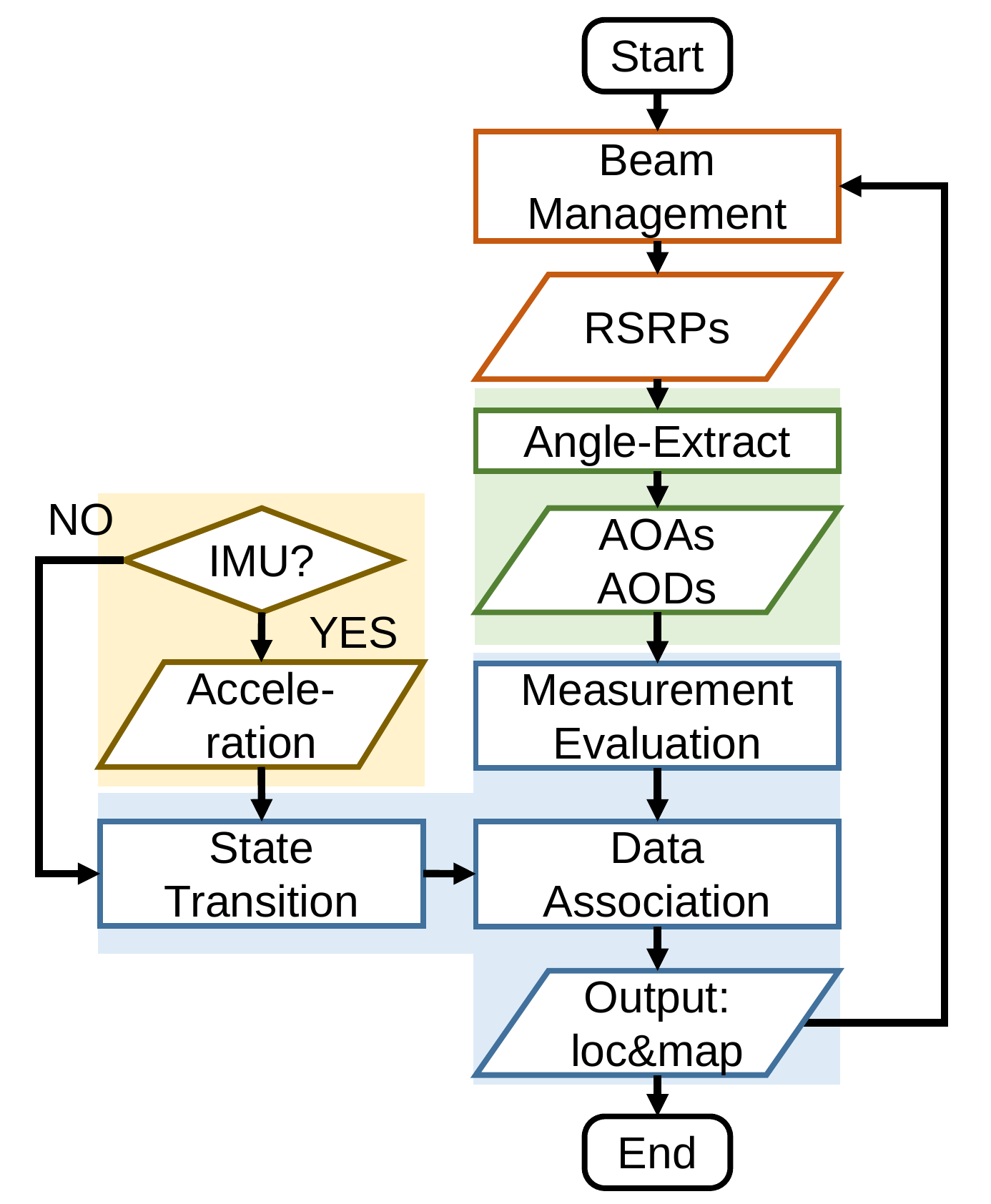}
	\caption{Flowchart of the angle-based SLAM algorithm, where the green, blue, and yellow blocks correspond to Angle-Extract, Angle-SLAM, and IMU-Calib parts, respectively.} \label{fig:fc}
\end{figure}

We obtain RSRP measurements using the mmWave hardware platform described in Section \ref{HP}. In this section, we propose an angle-based SLAM algorithm to track the UE and map features in the radio environment, as shown in Fig. \ref{fig:fc}. The algorithm consists of the following three parts.
\subsubsection{Angle-Extract}
We extract angle information from real RSRP measurements using a successive cancellation-based angle extraction method.
\subsubsection{Angle-SLAM}
We realize radio-based BP SLAM using only angle measurements.
\subsubsection{IMU-Calib}
We calibrate the estimation of the UE's track by embedding IMU measurements into the radio-based SLAM.

\begin{figure}
	\centering%
	\includegraphics[scale=0.48]{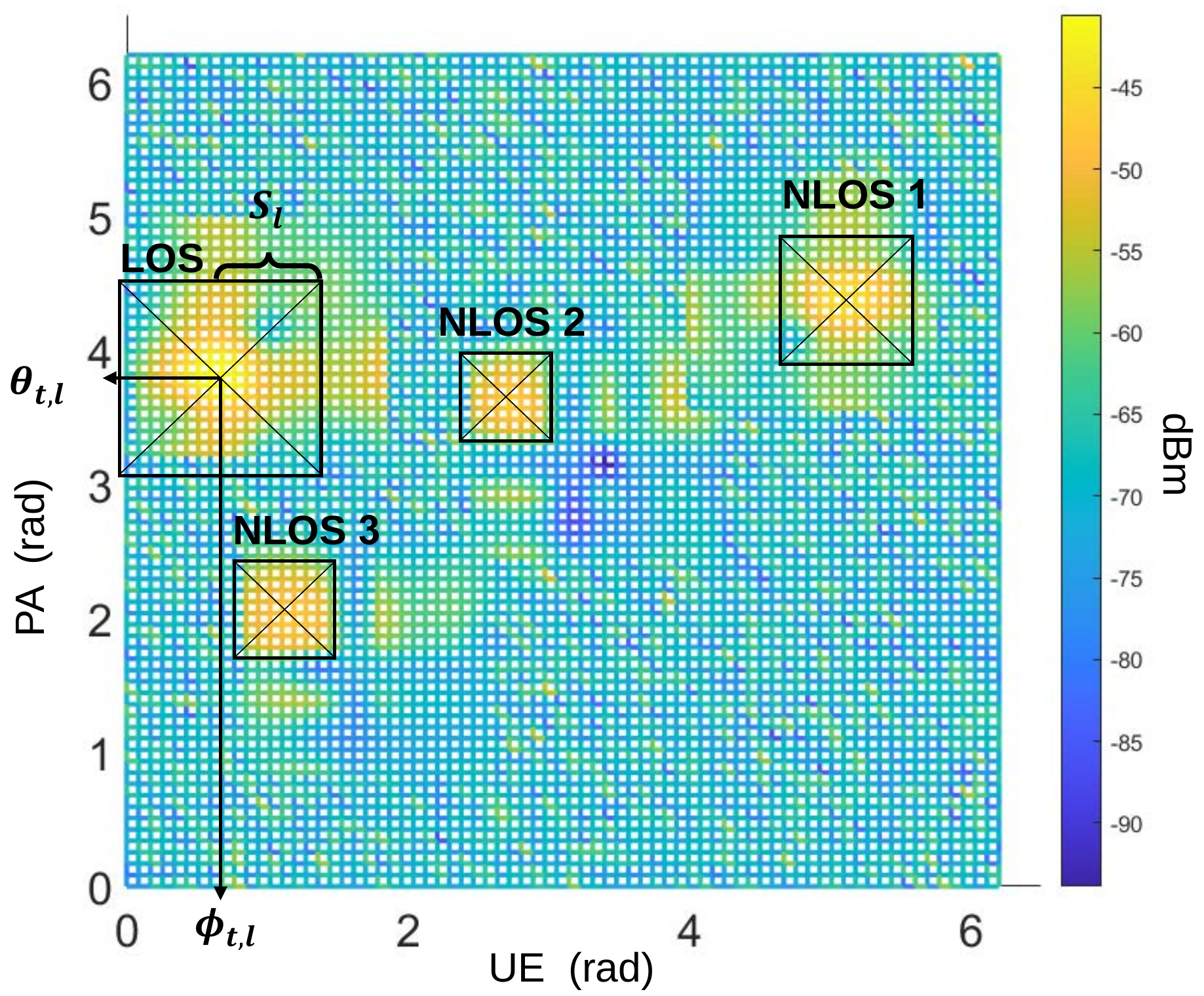}
	\caption{Visualization example of $\mathbf{R}$, where four paths are observed. The maximum RSRPs of LOS and NLOS 1, 2, and 3 are $-40.5$, $-44.5$, $-47.8$, and $-48.8$ dBm, respectively.} \label{fig:rsrp}\vspace{-0.3cm}
\end{figure}

\subsection{Angle-Extract}\label{sc}

We set three orientations of the UE and PA to cover $360^{\circ}$ in total, given that one mmWave phased array covers approximately $120^{\circ}$. The beam sweeping is executed following Section \ref{sweeping}. We obtain $64$ RSRPs recorded as 
${\mathbf{r}_{p,q} =[r_{0,0} ,\ldots,r_{0,7},r_{1,0} ,\ldots r_{1,7},\ldots, r_{7,0} ,\ldots,r_{7,7} ]}$ for orientation pair $(p,q)$, where $p$ and ${q=1,2,3}$.
We reshape the vector $\mathbf{r}_{p,q}$ to an RSRP matrix as
\begin{equation}\label{RSRP}
\mathbf{R}_{p,q}=
\begin{bmatrix}
r_{5,5}&r_{6,5}&r_{7,5}&r_{0,5}&r_{1,5}&r_{2,5}&r_{3,5}&r_{4,5}\\
r_{5,6}&r_{6,6}&r_{7,6}&r_{0,6}&r_{1,6}&r_{2,6}&r_{3,6}&r_{4,6}\\
r_{5,7}&r_{6,7}&r_{7,7}&r_{0,7}&r_{1,7}&r_{2,7}&r_{3,7}&r_{4,7}\\
r_{5,0}&r_{6,0}&r_{7,0}&r_{0,0}&r_{1,0}&r_{2,0}&r_{3,0}&r_{4,0}\\
r_{5,1}&r_{6,1}&r_{7,1}&r_{0,1}&r_{1,1}&r_{2,1}&r_{3,1}&r_{4,1}\\
r_{5,2}&r_{6,2}&r_{7,2}&r_{0,2}&r_{1,2}&r_{2,2}&r_{3,2}&r_{4,2}\\
r_{5,3}&r_{6,3}&r_{7,3}&r_{0,3}&r_{1,3}&r_{2,3}&r_{3,3}&r_{4,3}\\
r_{5,4}&r_{6,4}&r_{7,4}&r_{0,4}&r_{1,4}&r_{2,4}&r_{3,4}&r_{4,4}
\end{bmatrix}.
\end{equation} 
After orientation compensation, the column and row indexes in \eqref{RSRP} correspond to the same beam directions as those in \eqref{book}. By grouping all orientation pairs together, we obtain  
$$\mathbf{R}=
\begin{bmatrix}
\mathbf{R}_{1,1}&\mathbf{R}_{1,2}&\mathbf{R}_{1,3}\\
\mathbf{R}_{2,1}&\mathbf{R}_{2,2}&\mathbf{R}_{2,3}\\
\mathbf{R}_{3,1}&\mathbf{R}_{3,2}&\mathbf{R}_{3,3}
\end{bmatrix}.$$
$\mathbf{R}$ varies over time, but we remove the time index to provide a concise notation. 
We visualize a realization of $\mathbf{R}$ using nonlinear interpolation, as shown in Fig. \ref{fig:rsrp}, where four paths are observed. The result is obtained using real measurements from the hardware platform described in Section \ref{HP} and the experimental scenario explained in Section \ref{S1}. The maximum RSRPs of LOS and NLOS 1, 2, and 3 are $-40.5$, $-44.5$, $-47.8$, and $-48.8$ dBm, respectively. The energy gap between LOS and NLOS paths is $4$-$8$ dBm. Each path corresponds to a support in the matrix $\mathbf{R}$ due to energy spread. The index of the strongest element in the support corresponds to the angle information of that path. This observation motivates the application of a successive cancellation-based method to extract angle measurements from the grouped RSRP matrix $\mathbf{R}$.

We illustrate the detection of the first path using an example. First, we detect the index of the strongest element in matrix $\mathbf{R}$, denoted as $(r_1,c_1)$. After orientation compensation, we look up the angle dictionary in Equation \eqref{book} to find $\hat{\theta}_{t,1}$ and $\hat{\phi}_{t,1}$ that correspond to $(r_1,c_1)$. We then determine the support box of the first path centered at $(r_1,c_1)$. Let $S_1$ denote the half length of the support box, and initialize $S_1=1$. We define $\tilde{\eta}_1$ and $\eta_1$ as follows: 
\begin{equation}
\tilde{\eta}_1 = \sum\limits_{i=r_1-S_1-1}^{r_1+S_1+1}\sum\limits_{j=c_1-S_1-1}^{c_1+S_1+1}{\mathbf{R}}(i,j),
\end{equation}
and
\begin{equation}
\eta_1 = \sum\limits_{i=r_1-S_1}^{r_1+S_1}\sum\limits_{j=c_1-S_1}^{c_1+S_1}{\mathbf{R}}(i,j).
\end{equation}
If ${(\tilde{\eta}_1- \eta_1)}/{\eta_1}>\epsilon_{\rm se}$, where $\epsilon_{\rm se}$ denotes the rate threshold of the increased energy in a support, we increase the support box by letting $S_1=S_1+1$.
If the proportion of the increased energy is smaller than $\epsilon_{\rm se}$, we believe that the support can cover the current path. We then let $\tilde{\mathbf{R}}$ denote the residual RSRP measurement matrix and remove the influence of the first path using the following equation:
\begin{equation}
\tilde{\mathbf{R}}=\mathbf{R}- {\mathbf{M}}_1 \odot \mathbf{R},
\end{equation}
where ${\mathbf{M}}_1$ denotes the support of the first path, and 
\begin{equation}\label{mask}
{\mathbf{M}}_1(i,j)=\left\{
\begin{array}{ll}
1, & \text{ $r_1-S_1\leqslant i\leqslant r_1+S_1$} \\  & \text{ $c_1-S_1\leqslant j\leqslant c_1+S_1$},\\
0,& \text{otherwise},
\end{array} \right.
\end{equation}
where $\odot$ represents the Hadamard product. Let $\epsilon_{\rm re}$ denote the threshold of the remaining energy. If the remaining energy of $\tilde{\mathbf{R}}$ is larger than $\epsilon_{\rm re}$, we repeat the detection process. The algorithm terminates when ${\sum_{i}\sum_{j}\tilde{\mathbf{R}}(i,j)\leqslant\epsilon_{\rm re}}$. We then determine $\hat{L}_{t}$. Given that the SLAM algorithm introduced in Section \ref{bp} can deal with miss detection and false alarm, the slight estimation error of $ \hat {L}_{t} $ is acceptable. Therefore, using the proposed angle extraction method, we can obtain $\mathbf{z}_{t}=[(\hat{\theta}_{t,1},\hat{\phi}_{t,1}),\ldots,(\hat{\theta}_{t,\hat{L}_{t}},\hat{\phi}_{t,\hat{L}_{t}})]$. The proposed method is summarized in Algorithm \ref{alg1}.

\begin{algorithm}	
	\caption{\textbf{: Pseudocode of Angle-Extract}}\label{alg1}	
	\begin{algorithmic}[1]	
		\Require $\tilde{\mathbf{R}} \leftarrow\mathbf{R}$, $l \leftarrow 0$.
		\Ensure 	$\hat{\theta}_{t,l}$ and $\hat{\phi}_{t,l}$ for $l=1,\ldots,\hat{L}_{t}$.
		\While{$\sum_{i}\sum_{j}\tilde{\mathbf{R}}(i,j)>\epsilon_{\rm re}$}
		\State $l \leftarrow l+1$.
		\State Detect the index of the strongest element in $\tilde{\mathbf{R}}$, denoted as $(r_l,c_l)$.
		\State Lookup the angle dictionary
		\eqref{book} to find $\hat{\theta}_{t,l}$ and $\hat{\phi}_{t,l}$ that corresponding to $(r_l,c_l)$.
		\State $S_l\leftarrow1$.
		\While{${(\tilde{\eta}_l- \eta_l)}/{\eta_l}>\epsilon_{\rm se},$}		
		\State $S_l\leftarrow S_l+1$.
		\EndWhile		
		\State  Remove the $l$-th path by  $\tilde{\mathbf{R}}\leftarrow\tilde{\mathbf{R}}-{\mathbf{M}}_l \odot \tilde{\mathbf{R}}$.
		\EndWhile
	\end{algorithmic}
\end{algorithm}

\subsection{Angle-SLAM}\label{bp}
The goal of SLAM is to determine the state and location of PAs, VAs, RSPs, and tracks of UEs using the obtained angle measurements. In this subsection, we illustrate this approach using a single PA and a single UE. As UEs and PAs can operate in TDD mode, the SLAM mechanism described in this section can be readily extended to scenarios involving multiple PAs and UEs \cite{yj5}.

We consider PAs and VAs as features (landmarks) of the radio environment because $\mathbf{p}_{{\rm rsp},t,l}$ changes when the UE moves, while PAs and VAs have stable locations. We define ${\mathbf{p}_{t,1} = \mathbf{p}_{{\rm pa},t}}$ and ${\mathbf{p}_{t,l} = \mathbf{p}_{{\rm va},t,l}}$ for ${l\geqslant 2}$. Therefore, the location of the $l$-th feature is denoted by $\mathbf{p}_{t,l}$.
Due to data association uncertainty, a measurement can originate from a {\bf legacy feature} or a {\bf new feature}, or it may not originate from any feature (i.e., a {\bf false alarm}). A legacy feature means the feature already exists at time $t-1$, while a new feature means the feature does not exist at time $t-1$ but appears at time $t$. We use $\tilde{{\mathbf{v}}}_{t}$ and $\breve{{\mathbf{v}}}_{t}$ to denote the state of legacy and new features, respectively. We have ${\tilde{\mathbf{v}}_{t,l} = [\tilde{\mathbf{p}}_{t,l},\tilde{r}_{t,l}]}$ for ${l=1,\ldots, |\mathcal{D}_{t}|}$, where $\tilde{\mathbf{p}}_{t,l}$ denotes the location of a legacy feature and $\mathcal{D}_{t}$ denotes the set of legacy feature indexes that generate a measurement at time $t$. Similarly, we have ${\breve{\mathbf{v}}_{t,l} = [\breve{\mathbf{p}}_{t,l},\breve{r}_{t,l}]}$ for ${l=1,\ldots,|\mathcal{N}_{t}|}$, where
$\breve{\mathbf{p}}_{t,l}$ denotes the location of a new feature and $\mathcal{N}_{t}$ denotes a set of measurement indexes originating from new features. Binary variables ${\tilde{r}_{t,l}\in\{0,1\}}$ and
${\breve{r}_{t,l}\in\{0,1\}}$ indicate the existence of the $l$-th feature at time $t$; that is, the feature exists at time $t$ if and only if ${\tilde{r}_{t,l}=1}$ or ${\breve{r}_{t,l}=1}$. We denote
${{\mathbf{v}}_{t}=[\tilde{{\mathbf{v}}}_{t},\breve{{\mathbf{v}}}_{t}]}$. Moreover, $\mathcal{F}_{t}$ denote a set of measurement indexes of false alarms. Therefore, we classify the measurement indexes in $\mathcal{M}_{t}$ into three subsets according to their origins and obtain
$
|\mathcal{M}_{t}| = |\mathcal{D}_{t}|+|\mathcal{N}_{t}|+|\mathcal{F}_{t}|
$.

The joint posterior probability density function (PDF) of the state of the UE and features and the data association vectors conditioned on measurements for all times up to $T$ is defined as 
\begin{multline}\label{jointP}
f(\mathbf{u}_{1:T},{\mathbf{v}}_{1:T}, \mathbf{p}_{{\rm rsp},1:T}, \bm{\psi}_{1:T}|\mathbf{z}_{1:T}) \\
= \prod\limits_{t=1}^{T} f(\mathbf{u}_{t},{\mathbf{v}}_{t},\mathbf{p}_{{\rm rsp},t},\bm{\psi}_{t}|\mathbf{z}_{t}),
\end{multline}
where the data association vector is denoted as $\bm{\psi}$ and is comprehensively explained in Appendix \ref{A}. 
Using Bayes' theorem, we have 
\begin{multline}\label{jointP1}
f(\mathbf{u}_{1:T},{\mathbf{v}}_{1:T}, \mathbf{p}_{{\rm rsp},1:T}, \bm{\psi}_{1:T}|\mathbf{z}_{1:T})\\
\!\propto \! \prod\limits_{t=1}^{T}\underbrace {f(\mathbf{u}_{t},\tilde{{\mathbf{v}}}_{t}|\mathbf{u}_{t-1},{\mathbf{v}}_{t})}_{(a)}
\underbrace {f(\mathbf{z}_{t}|\mathbf{u}_{t},{\mathbf{v}}_{t},\mathbf{p}_{{\rm rsp},t}, \bm{\psi}_{t})}_{(b)}\\
\times\underbrace {f( \bm{\psi}_{t},{c}_{t},\breve{{\mathbf{v}}}_{t},\mathbf{p}_{{\rm rsp},t}|\tilde{{\mathbf{v}}}_{t},\mathbf{u}_{t})}_{(c)},
\end{multline}
where ${{c}_{t}=|\mathcal{M}_{t}|}$ is the number of measurements at time $t$. Moreover, (a), (b), and (c) in \eqref{jointP1} correspond to the state transition, measurement evaluation, and data association phases, respectively. The data fusion phase corresponds to the entire process of \eqref{jointP1}. The detailed derivations of (a), (b), and (c) in \eqref{jointP1} are given in Appendix \ref{B}.

\begin{figure}
	\centering
	\vspace{-0.6cm}
	\includegraphics[scale=0.4]{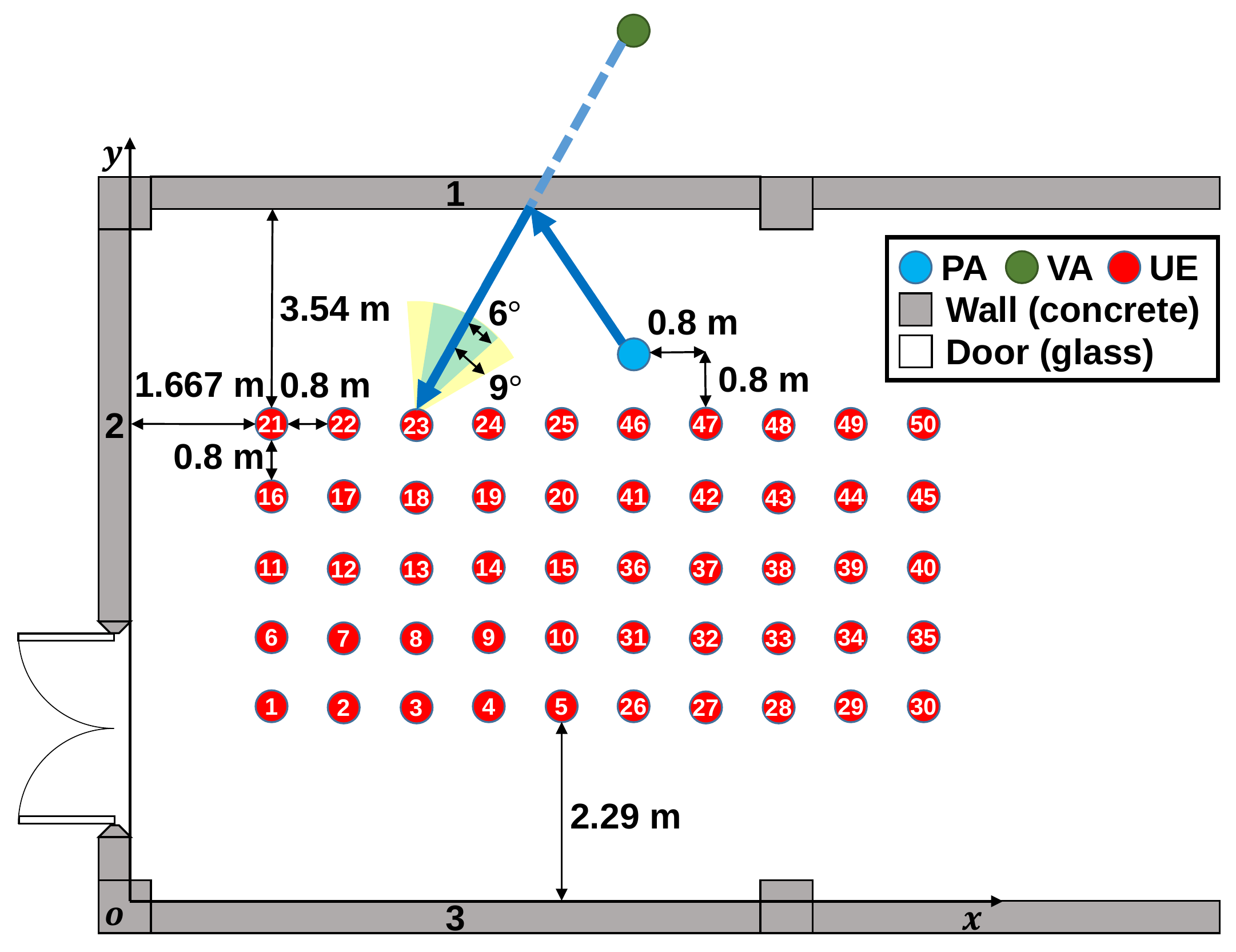}
	\caption{Floor plan of the empty hall in scenario $1$, where three walls are numbered, and the VA corresponding to wall $1$ is depicted. Yellow and green sectors indicate the difference in measurement and specular reflection angles, which are smaller than $9^{\circ}$ and $6^{\circ}$, respectively.}
	\label{fig:S1}\vspace{-0.3cm}
\end{figure}

A minimum mean squared error (MMSE) estimator for the UE's state $\mathbf{u}_{T} $ at time $T$ is given as
\begin{equation}\label{mmse1}
\hat{\mathbf{u}}_{T} = \int \mathbf{u}_{T}  f(\mathbf{u}_{T} |\mathbf{z}_{1:T})  \text{d} \mathbf{u}_{T},
\end{equation}
where $f(\mathbf{u}_{T}|\mathbf{z}_{1:T})= \int_{\mathbf{x}}  f(\mathbf{u}_{1:T},{\mathbf{v}}_{1:T},\mathbf{p}_{{\rm rsp},1:T}, \bm{\psi}_{1:T}|\mathbf{z}_{1:T}) \text{d}\mathbf{x} $ and ${\mathbf{x} = [\mathbf{u}_{1:T\!-\!1},\!{\mathbf{v}}_{1:T},\mathbf{p}_{{\rm rsp},1:T},\! \bm{\psi}_{1:T}]}$.
The posterior existence probability is given by
\begin{equation}\label{marginal}
p({r}_{T,l}\!=\!1|\mathbf{z}_{1:T}) =\int \!\!\!f(\mathbf{{p}}_{T,l},\!{r}_{T,l}\!=\!1|\mathbf{z}_{1:T}) \text{d} \mathbf{p}_{T,l},
\end{equation}
where $f(\mathbf{{p}}_{T,l},\!{r}_{T,l}\!=\!1|\mathbf{z}_{1:T})$ is a marginal posterior PDF in \eqref{jointP1}.
On the basis of Bayes' theorem, we obtain
\begin{equation}\label{marginal2}
f(\mathbf{p}_{T,l}|{r}_{T,l}\!=\!1,\mathbf{z}_{1:T})  \!=\! \frac{ f(\mathbf{p}_{T,l},{r}_{T,l}\!=\!1|\mathbf{z}_{1:T})}{p({r}_{T,l}=1|\mathbf{z}_{1:T})}.
\end{equation}
The MMSE estimator for the feature $\mathbf{p}_{T,l}$ can be obtained as
\begin{equation}\label{mmse6}
\hat{\mathbf{p}}_{T,l} \! =\!\! \int \!\! \mathbf{p}_{T,l} f(\mathbf{p}_{T,l}|{r}_{T,l}=1,\mathbf{z}_{1:T})  \text{d} \mathbf{p}_{T,l}.
\end{equation}
The detection phase is \eqref{marginal}, and the estimation phase is \eqref{mmse1} and \eqref{mmse6}. 
However, direct marginalization from \eqref{jointP1} is infeasible; hence, we use particle-based implementation \cite{rs4} to approximate the continuous messages.

\begin{figure}
	\centering
	\vspace{-0.1cm}
	\includegraphics[scale=0.61]{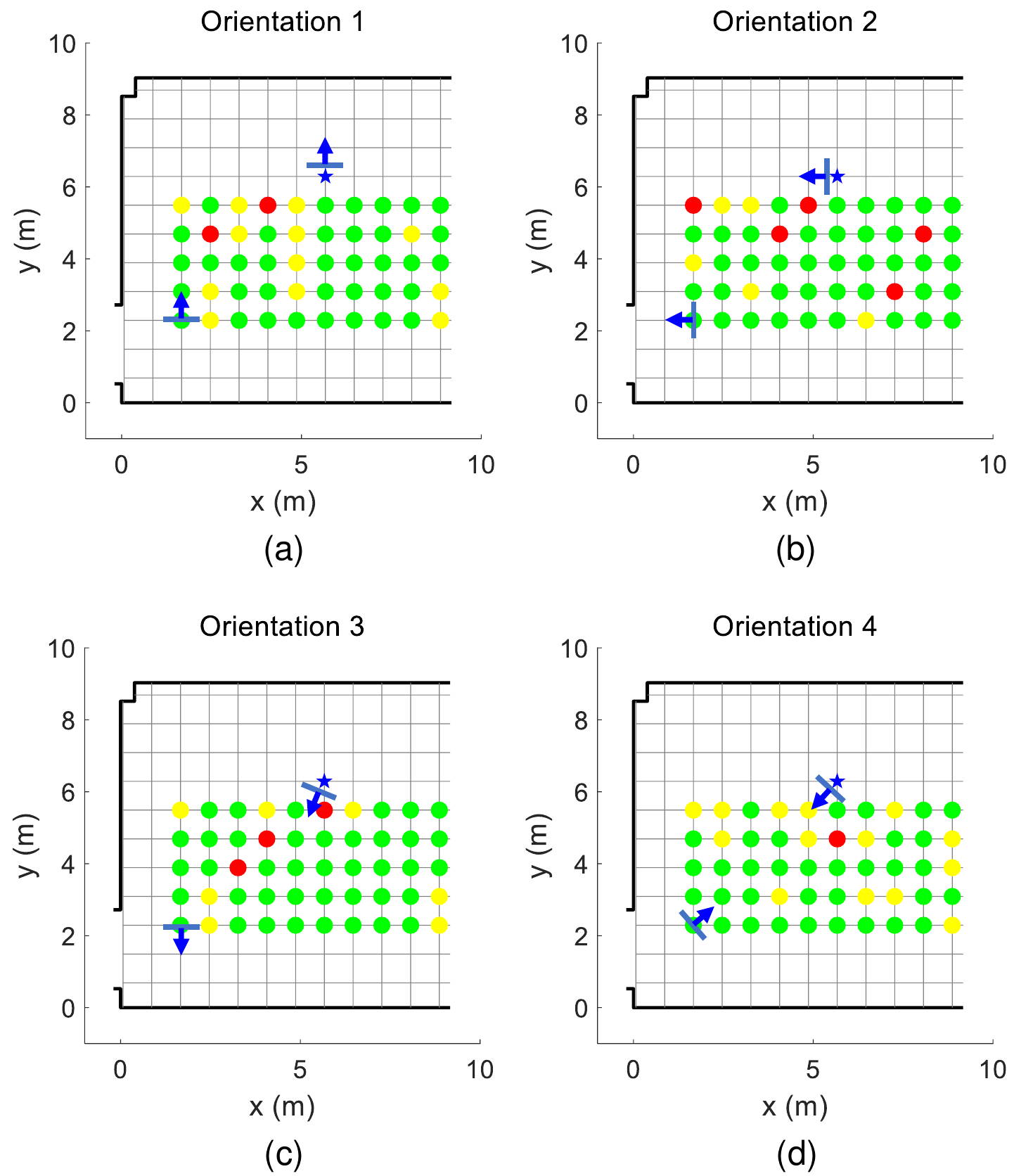}
	\caption{Comparison of angle measurements with theoretical specular reflections. The blue arrows illustrate the UE and PA orientations of the first 25 UE points. Green, yellow, and red represent high, medium, and low match levels, respectively.
	} \label{fig:BA}\vspace{-0.3cm}
\end{figure}

\begin{figure*}
	\centering
	\includegraphics[scale=0.62]{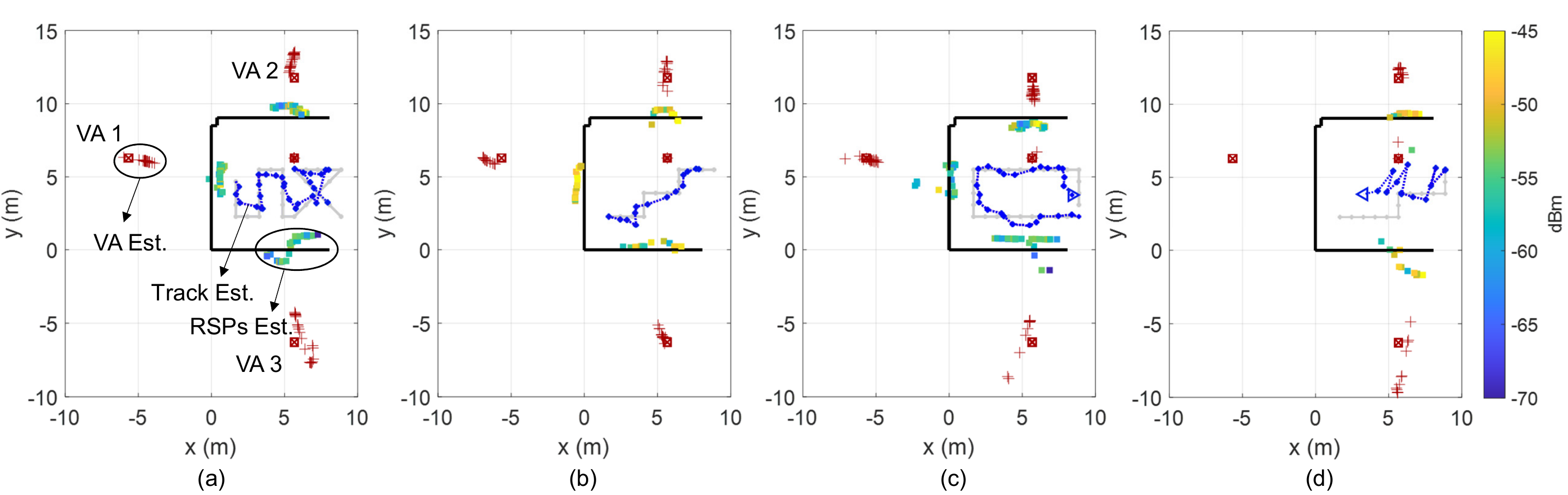}
	\caption{SLAM performance of four different tracks, where the red crossed box, gray solid line, and black solid lines denote the true PA/VAs, track, and walls, respectively; the red crosses, blue dotted line, and points in heat map colors are estimated PA/VAs, track, and RSPs, respectively. (a) Track-a; (b) Track-b; (c) Track-c;
			(d) Track-d.} \label{fig:S1A1SLAM}
\end{figure*}
\begin{figure*}
	\centering
	\includegraphics[scale=0.62]{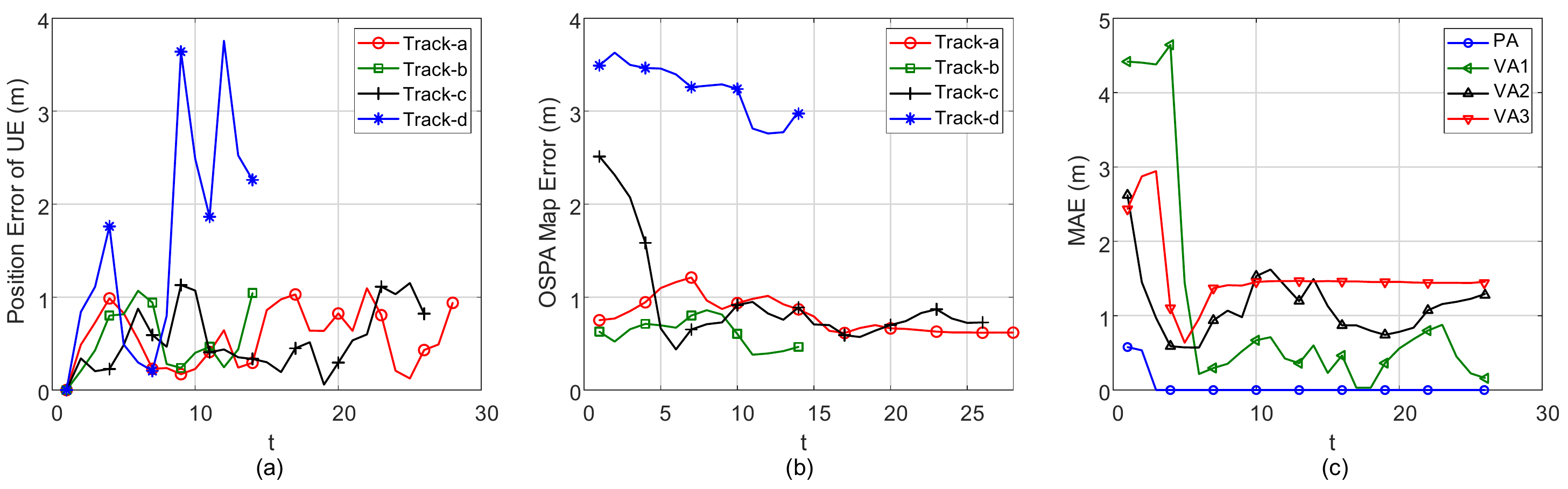}
	\caption{(a) Localization performance of different tracks. (b) Mapping performance of different tracks. (c) Mapping performance of different VAs for Track-c.} \label{fig:S1A1LM}
\end{figure*}

\begin{table}
	\centering
	\small
	\caption{Array orientations for UE and PA in scenario 1}\label{T1}
	\begin{tabular}{ccccccccc}
		\toprule  
		&\multicolumn{1}{c}{O1}&\multicolumn{1}{c}{O2}&\multicolumn{2}{c}{O3}&\multicolumn{2}{c}{O4}\\
		&$\alpha_{{\rm u}}$($\alpha_{{\rm pa}}$)&$\alpha_{{\rm u}}$($\alpha_{{\rm pa}}$)&$\alpha_{{\rm u}}$&$\alpha_{{\rm pa}}$&$\alpha_{{\rm u}}$&$\alpha_{{\rm pa}}$ \\
		\midrule  
		$1-25$&$ 0^\circ$&$ 90^\circ$&$ 180^\circ$&$ 153.4^\circ$&$ -45^\circ$&$ 135^\circ$ \\
		$26-50$&$ 0^\circ$&$ 90^\circ$&$ 180^\circ$&$ 180^\circ$&$ 45^\circ$&$ 225^\circ$\\
		\bottomrule  
	\end{tabular}
\end{table}

\subsection{IMU-Calib}\label{imu}

The states of UE and legacy features are assumed to evolve and follow Markovian state dynamics independently. If no IMU (inertial measurement unit) information is available, we assume that the state transition function $f(\mathbf{u}_{t}|\mathbf{u}_{t-1})$ of the UE is defined by a linear, near-constant velocity motion model \cite{sem}: 
\begin{equation}\label{ncv}
\mathbf{u}_{t}^{\rm T} = \mathbf{A} \mathbf{u}_{t-1}^{\rm T}+ \bm{\omega}_{t},
\end{equation}
where
\begin{equation}
\mathbf{A}=\left(
\begin{array}{cccc}
1 & 0 & \Delta T & 0\\
0 & 1 & 0 & \Delta T\\
0 & 0 & 1 & 0\\
0 & 0 & 0 & 1\\
\end{array}
\right),
\end{equation}
$\Delta T$ is the beam sweeping period, and $\bm{\omega}_{t}$ is the driving process that follows an independently identical Gaussian distribution across $t$ with zero mean. 
For tracks where the UE turns frequently, the model in \eqref{ncv} cannot accurately represent the motion of the UE.

\begin{figure*}
	\includegraphics[scale=0.62]{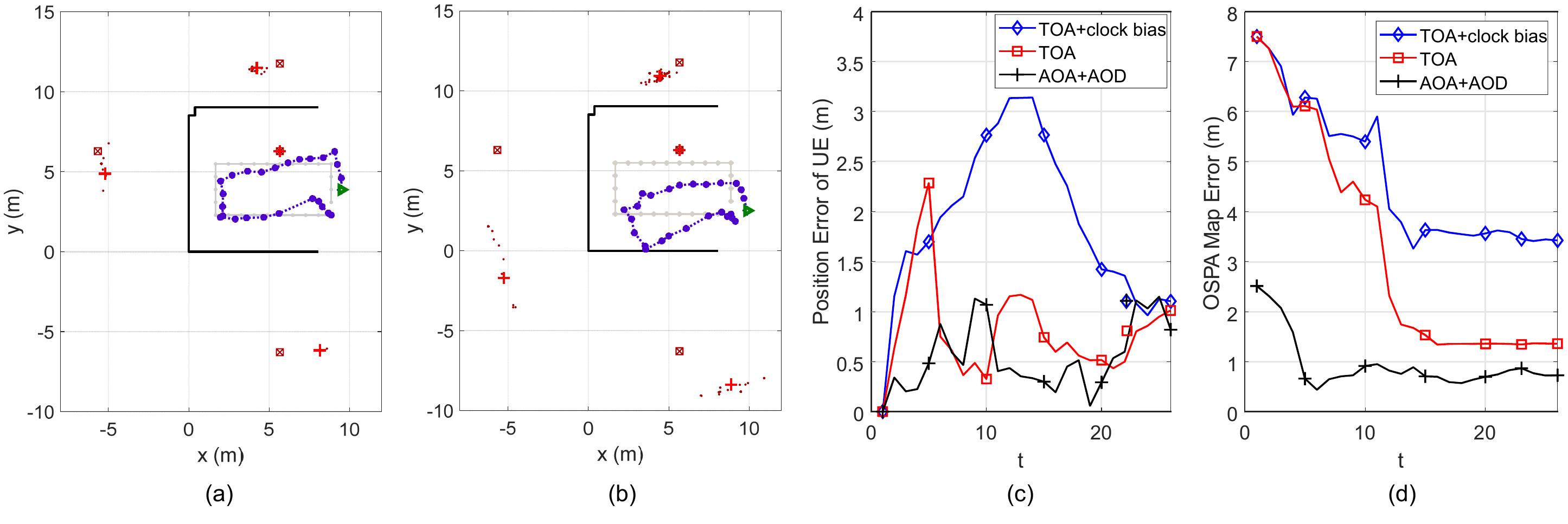}
	\caption{Performance comparison of different SLAM schemes: (a) TOA-based SLAM with $0.3$ ns standard deviation; (b) TOA-based SLAM with $0.3$ ns standard deviation and $3$ ns clock bias; (c) Localization accuracy comparison of the TOA- and proposed angle-based SLAM schemes; (d) Mapping performance comparison of the TOA- and proposed angle-based SLAM schemes.} \label{fig:TOA}
\end{figure*}

\begin{figure*}
	\centering
	\includegraphics[scale=0.62]{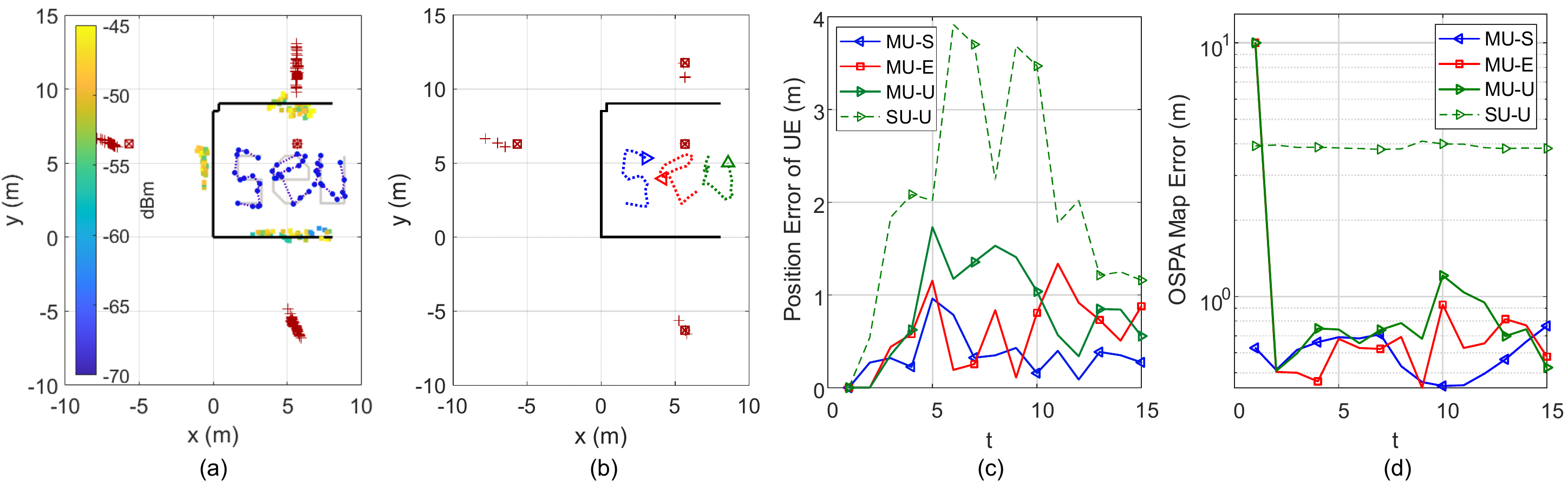}
	\caption{(a) SLAM performance of crowdsourcing mechanism, where Track-s, Track-e, and Track-u enter the ROI at $t=1$, $t=2$, and $t=2$, respectively. (b) Letter display of the estimated tracks, where blue, red, and green triangles denote the endpoints of three tracks. (c) Localization performance of crowdsourcing mechanism. (d) Mapping performance of crowdsourcing mechanism.} \label{fig:S1A2SLAM}
\end{figure*}

If IMU measurements are available, these measurements can be fused into the state transition module of the BP SLAM algorithm. In this study, we consider the acceleration of the UE measured by the equipped IMU, denoted as $\bm{\varsigma}_t=[a_{{\rm x},t},a_{{\rm y},t}]$.\footnote{Our study is easily extended to 3D scenarios, where $\mathbf{u}$, $\mathbf{\varsigma}$, $\mathbf{A}$, and $\mathbf{B}$ will be a $6$-D vector, $3$-D vector, $6\times 6$ matrix, and $6\times 3$ matrix, respectively.} Then, we have 
\begin{equation}\label{vv}
\mathbf{u}_{t}^{\rm T} = \mathbf{A} \mathbf{u}_{t-1}^{\rm T} +\mathbf{B}\bm{\varsigma}_{t-1}^{\rm T} +\bm{\omega}_{t},
\end{equation}
where
\begin{equation}
\mathbf{B}=\left(
\begin{array}{cccc}
\frac{1}{2}\Delta T^2 & 0 \\
0 & \frac{1}{2}\Delta T^2  \\
 \Delta T & 0\\
 0 & \Delta T\\
\end{array}
\right).
\end{equation}
Noise caused by accelerometer bias is denoted as $\mathbf{\omega}_t$. We use the same notation $\mathbf{\omega}_t$ for driving noise in \eqref{ncv} and accelerometer noise in \eqref{vv} to simplify the representation. Acceleration can describe changes in the direction and magnitude of velocity. Therefore, the model in \eqref{vv}, which is calibrated by the IMU, can approach the ground truth.

\section{Experimental Results}	\label{result}
We conduct extensive experiments to verify the feasibility of the proposed angle-based SLAM with the implemented mmWave ISAC platform. In this section, we aim to verify the following points: 
\begin{itemize}
	\item Do mmWaves experience specular reflection on smooth walls?
	\item Can multiuser collaboration improve sensing performance?
	\item How does sensing performance improve with the assistance of an IMU?
	\item How does sensing performance vary for nonsmooth surfaces?
	\item How does sensing performance vary in scenarios with the birth and death of beams?
\end{itemize}

\subsection{Scenario 1: Empty Hall}\label{S1}

\begin{figure*}
	\centering
	\includegraphics[scale=0.55]{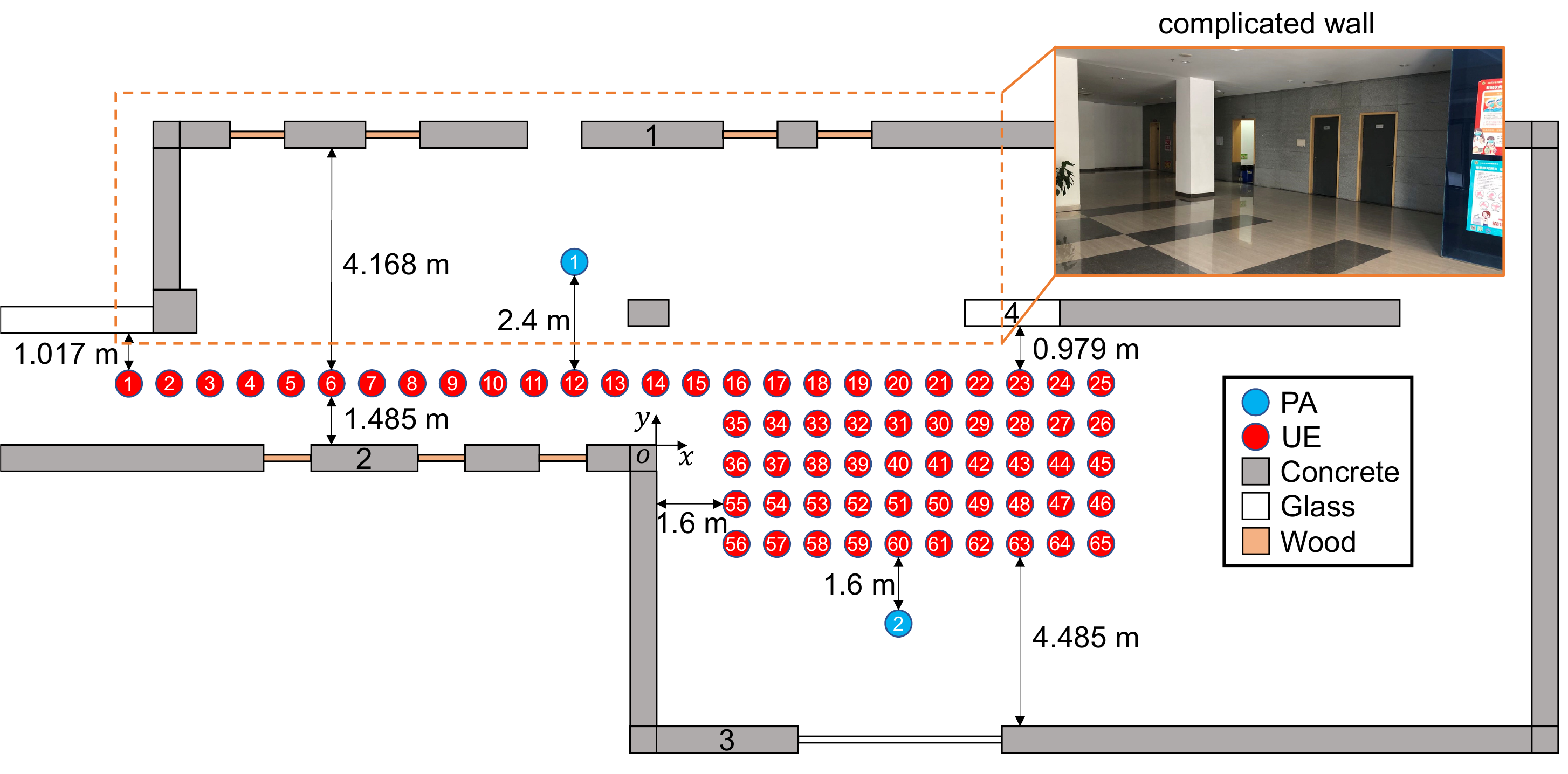}
	\caption{Floor plan of scenario $2$, where four walls are numbered.} \label{fig:S2}
\end{figure*}

\subsubsection{Settings}
We consider an empty indoor hall, as shown in Fig. \ref{fig:S1}. The PA is located at coordinates $(5.667,\ 6.290)$ m, and we illustrate the scenario with one VA located at $(5.667,\ 9.030)$ m. The UE moves within the red marked area, which we refer to as the UE active zone. The spacing between adjacent red dots in the horizontal or vertical direction is 0.8 m. We divide the UE active zone into 50 grid points to facilitate the description of its tracks.
The UE and PA perform beam sweeping according to the 5G NR beam management described in Section \ref{BM}. This scenario involves four main propagation paths: one LOS path and three specular NLOS paths from walls 1 to 3. We set four different array orientations for the UE and PA to sense the surrounding environment at each grid point in the UE active zone, as described in Table \ref{T1}.\footnote{For convenience, we set two different array orientations of the PA in Orientation 3 and of the UE and PA in Orientation 4 for the first 25 and later 25 grid points in the UE active zone.}
According to \cite{rs4,yj5,jy}, the parameters involved in the BP SLAM algorithm are as follows: the variance of the driving process $\bm{\omega}{t}$ is 0.0711; detection probability $P_{\rm d}=0.95$; survival probability $P_{\rm s}=0.999$; mean of false alarms $\mu_{\rm false}=0.1$; mean of newly born features $\mu_{\rm new}=10^{-6}$; step length is 0.8 m; unreliability threshold is $10^{-4}$; detection threshold is 0.5; and the number of particles is $10^5$.

\subsubsection{Specular Reflection}
In this subsection, we simplify the angle extraction process by selecting the maximum RSRP for the beam sweeping of each orientation pair and checking the angle dictionary for the corresponding AOA and AOD. The theoretical value of specular reflection at wall 1 is illustrated by the blue arrow in Fig. \ref{fig:S1}. We compare the measured angle of the propagation path with the theoretical value of specular reflection. If the angle difference is less than half of the minimum beam spacing (6$^\circ$ as depicted in Fig. \ref{fig:codebook}), we regard the measurement as highly matched with specular reflection. If the difference is more than 6$^\circ$ but less than half of the maximum beam spacing (9$^\circ$), we consider the measurement as moderately matched. If the difference is more than 9$^\circ$, we consider the measurements as mismatched. The effectiveness of matching the measurements with specular reflection is shown in Fig. \ref{fig:BA}, where green, yellow, and red represent high, medium, and low match levels, respectively. 

\textbf{Consequently, the reflection of electromagnetic waves at 28 GHz on the smooth concrete wall is consistent with specular reflection, which lays the foundation for the angle-based SLAM mechanism.}

\subsubsection{Single UE}
We selected four different tracks that start from the top left (Fig. \ref{fig:S1A1SLAM}(a)), bottom left (Fig. \ref{fig:S1A1SLAM}(b)), bottom right (Fig. \ref{fig:S1A1SLAM}(c)), and top right (Fig. \ref{fig:S1A1SLAM}(d)) of the UE active zone to evaluate the SLAM performance. The results are shown in Fig. \ref{fig:S1A1SLAM}, where the red box, gray solid line, and black solid line represent the true PA (VA), track, and wall, respectively. The red cross, blue dotted line, and dot in the heat map color represent the estimated PA (VA), track, and RSP, respectively.

The results in Figs. \ref{fig:S1A1LM}(a) and \ref{fig:S1A1LM}(b) show that the proposed mechanism for Tracks a, b, and c can achieve approximately 0.5 m mean position error and 0.6 m mean optimal subpattern assignment (OSPA) map error. We present the mean absolute error (MAE) of each estimated VA corresponding to Track-c in Fig. \ref{fig:S1A1LM}(c) to analyze the estimation error of each VA. The results reveal that the estimated VA 3 has the worst accuracy among all estimated VAs because it is the farthest from the PA. The accuracy of the estimated VA 1 increases as the UE moves closer to it. However, VA 1 cannot be sensed for Track-d because the AOA measurements of the LOS and NLOS paths from VA 1 are remarkably close to each other at the first few steps of Track-d, thereby causing data association errors.

Moreover, we compared the SLAM performance of the proposed scheme with the classic TOA-based BP SLAM algorithm \cite{rs1}. Achieving high-precision synchronization in real communication systems is difficult. Therefore, we used simulated TOA measurements to conduct the comparison. In the simulations, the standard deviation of TOA is 0.3 ns, and the corresponding standard deviation of distance is around 0.1 m. Figs. \ref{fig:TOA}(a) and \ref{fig:TOA}(b) show the SLAM result without clock bias and with 3 ns (1 m) clock bias, respectively. We also observed rotations of a random angle in Figs. \ref{fig:TOA}(a) and \ref{fig:TOA}(b). The localization and mapping performances of the TOA-based scheme were worse than the angle-based scheme according to Figs. \ref{fig:TOA}(c) and \ref{fig:TOA}(d). Notably, achieving even 0.3 ns standard deviation of TOA with 3 ns clock bias in currently used communication systems is difficult.

\textbf{The experimental results demonstrate that SLAM can be realized with communication signals during the beam sweeping process. Moreover, our mmWave ISAC platform can achieve decimeter-level positioning and mapping accuracy in most cases. However, the tracks of the UE also affect the SLAM results. This observation motivates this study to achieve satisfactory sensing performance of UEs with poor tracks through cooperation among multiple UEs.}

\begin{figure*}
	\centering
	\includegraphics[scale=0.59]{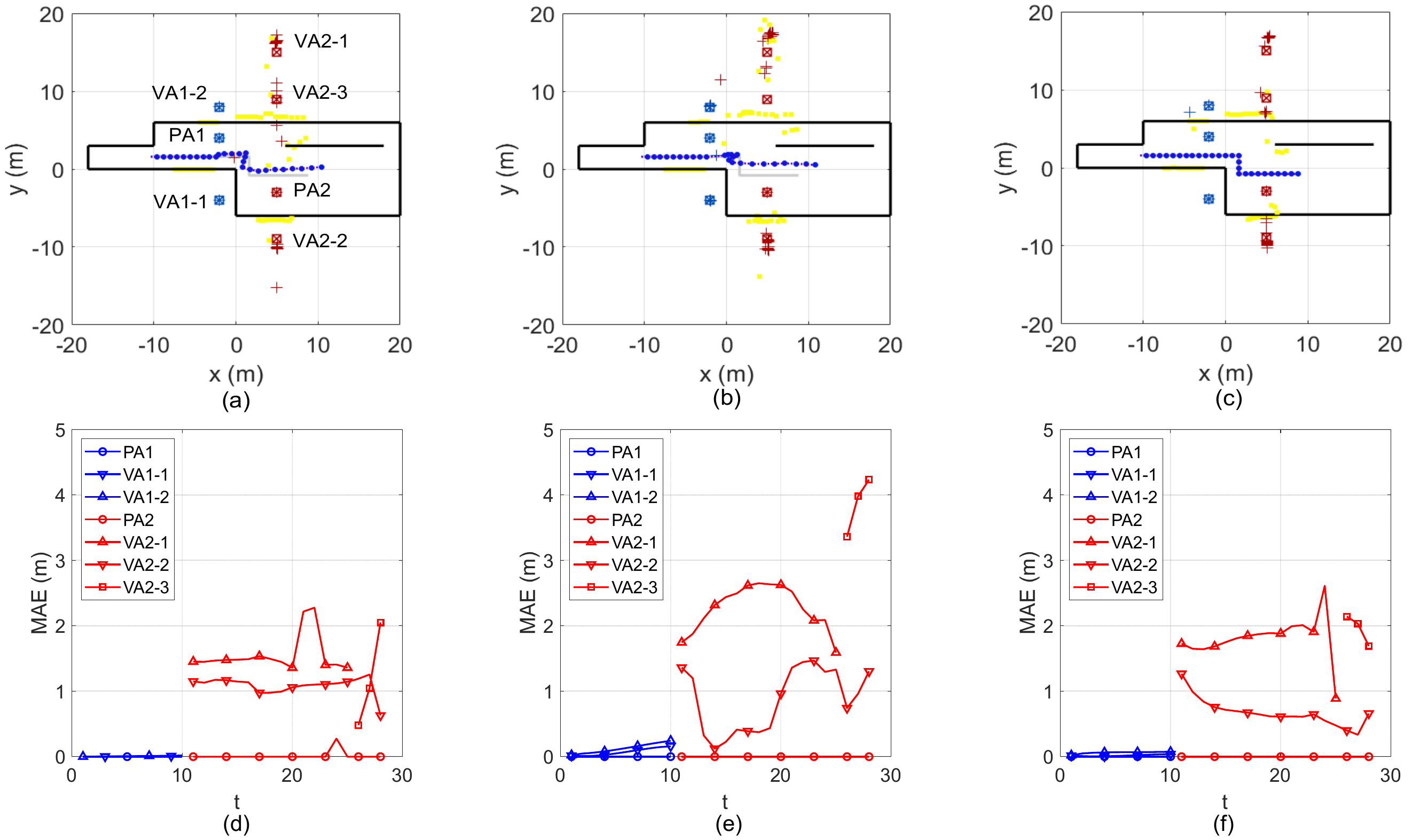}
	\caption{(a) SLAM performance for $1^{\circ}$ standard angle deviation.
		(b) SLAM performance for $6^{\circ}$ standard angle deviation.
		(c) SLAM performance for $6^{\circ}$ standard angle deviation with IMU assistance.
		(d) Mapping performance for $1^{\circ}$ standard angle deviation.
		(e) Mapping performance for $6^{\circ}$ standard angle deviation.		
		(f) Mapping performance for $6^{\circ}$ standard angle deviation with IMU assistance.} \label{fig:S2A1SLAM}
\end{figure*}

\subsubsection{Multiple UEs}

In this subsection, we adopt the crowdsourcing-based BP SLAM mechanism \cite{yj5}. We choose three tracks as shown in Fig. \ref{fig:S1A2SLAM}(a). Track-s enters the region of interest (ROI) at $t=1$, while Track-e and Track-u enter the ROI at $t=2$. Track-u starts from the top right of the UE active zone. Without crowdsourcing, the localization and mapping performance is poor, as shown in Figs. \ref{fig:S1A2SLAM}(c) and \ref{fig:S1A2SLAM}(d), where ``MU-U'' and ``SU-U'' denote the results with and without multiuser cooperation, respectively. As UEs of Track-e and Track-u enter the ROI at $t=2$, the position and OSPA map errors of MU-E and MU-U at $t=1$ are $0$ and $10$ m, respectively. With the crowdsourcing mechanism, the UEs of Track-e and Track-u inherit the estimated features of the radio environment from the first UE (Track-s). Therefore, all three UEs obtain their tracks and the environment features accurately, and the walls of the environment (comprising estimated RSPs) can be quickly constructed with rich tracks. Fig. \ref{fig:S1A2SLAM}(b) shows the letter display of the estimated tracks. 

\textbf{The experiment verifies the effectiveness of the crowdsourcing-based BP SLAM mechanism, where the localization and the OSPA mapping errors of Track-u are promoted by approximately $50 \%$ and $77 \%$, respectively.}

\begin{figure}
	\centering
	\includegraphics[scale=0.7]{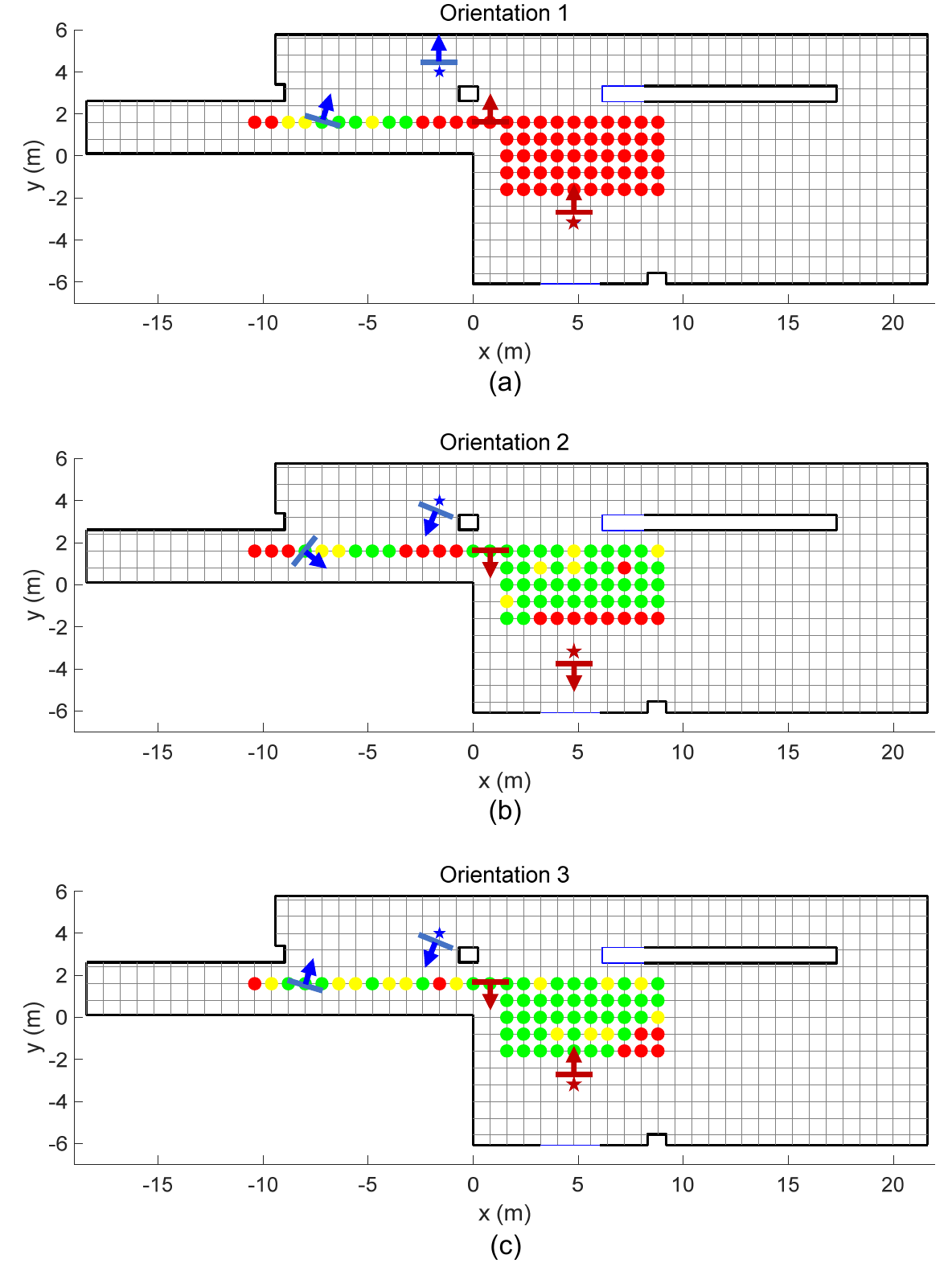}
	\caption{Comparison of angle measurements with theoretical specular reflections. Green, yellow, and red represent high, medium, and low match levels, respectively.} \label{fig:BA2}
\end{figure}

\begin{figure*}
	\centering
	\includegraphics[scale=0.622]{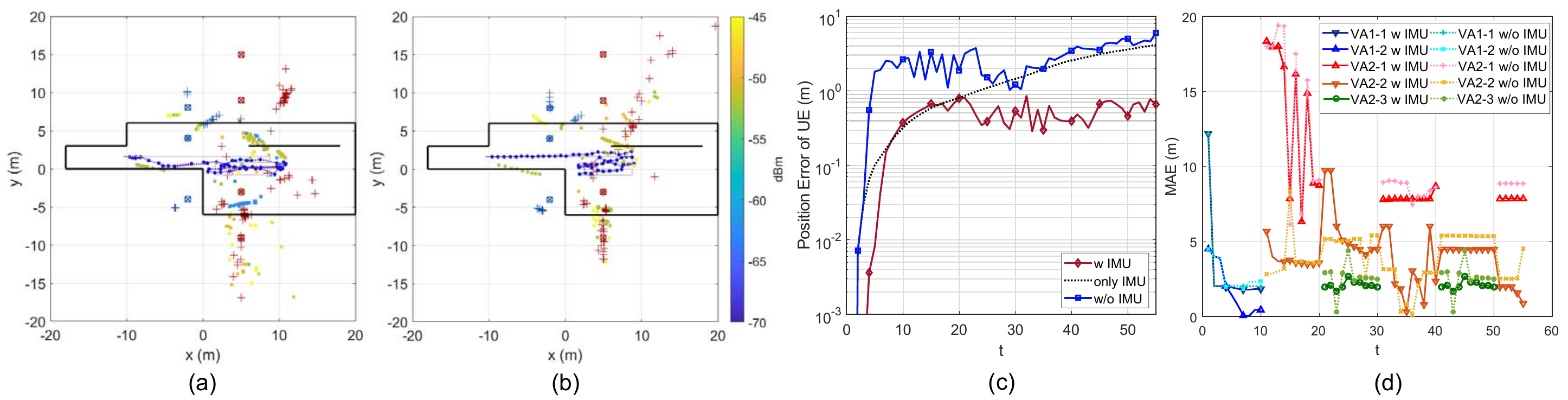}
	\caption{(a) SLAM performance when no IMU information is available. (b) SLAM performance with the assistance of IMU measurements. (c) Localization performance with and without IMU. (d) Mapping performance with and without IMU.} \label{fig:S2A2SLAM}
\end{figure*}

\begin{table}
	\centering
	\small
	\caption{Array orientations for UE and PA in scenario 2}\label{T2}
	\begin{tabular}{ccccccc}
		\toprule  
		&\multicolumn{2}{c}{O1}&\multicolumn{2}{c}{O2}&\multicolumn{2}{c}{O3}\\
		&$\alpha_{{\rm u}}$&$\alpha_{{\rm pa}}$&$\alpha_{{\rm u}}$&$\alpha_{{\rm pa}}$&$\alpha_{{\rm u}}$&$\alpha_{{\rm pa}}$ \\
		\midrule  
		$1\!-\!10$&$ -26.6^\circ$&$ 0^\circ$&$ 225^\circ$&$ 153.4^\circ$
		&$ -26.6^\circ$&$ 243.4^\circ$ \\
		$11\!-\!65$&$ 0^\circ$&$ 0^\circ$&$ 180^\circ$&$ 180^\circ$
		&$ 180^\circ$&$ 0^\circ$\\
		\bottomrule  
	\end{tabular}
\end{table}

\subsection{Scenario 2: Complex Corridor}
\subsubsection{Settings}
In this experiment, we consider a complex scenario with various obstacles, including corridors, wooden doors, glass decorative walls, and pillars. To add to the complexity, we introduce the phenomenon of beam birth and death, and we use two PAs located at different positions denoted as PA $1$ and PA $2$. The UE active zone is marked with $65$ grid points, where the UE can receive beams from PA $1$ for grid points $1$ to $10$ and from PA $2$ for grid points $11$ to $65$ (Fig. \ref{fig:S2}). Three different array orientations for UE and PA are used to sense the surrounding environment, as described in Table \ref{T2}.
The BP SLAM algorithm parameters are set as follows, based on \cite{rs4,yj5,jy}: the variance of the driving process $\bm{\omega}_{t}$ is $10^{-1}$ without IMU and $10^{-5}$ with IMU; the survival probability $P_{\rm s}=0.9$; the mean of false alarms $\mu_{\rm false}=0.05$; the mean of newly born features $\mu_{\rm new}=10^{-2}$; and the unreliability threshold is $10^{-2}$. The remaining parameter settings are the same as those in Section \ref{S1}.

\subsubsection{Simulations}
In this subsection, our aim is to explore the benefits of IMU in the proposed SLAM mechanism. To achieve this, we first analyze the SLAM performance in scenario 2 using raytracing-based simulations with three standard deviation settings of angle measurement: (a) $1^{\circ}$; (b) $6^{\circ}$; (c) $6^{\circ}$ with IMU assistance. The $6^{\circ}$ standard deviation is the smallest grid error of our designed beam codebook, and each wall is treated as a smooth reflective surface. The simulation results are presented in Fig. \ref{fig:S2A1SLAM}.
When the angle error is small, i.e., $1^{\circ}$, the UE can distinguish VA2-1 and VA2-3 (Fig. \ref{fig:S2A1SLAM}[a]). However, with a large angle error of $6^{\circ}$, errors are introduced in the data association of close VAs, such as VA2-1 and VA2-3, which leads to an increase in localization and mapping errors. However, when the IMU information is fused with the proposed SLAM mechanism, the UE can distinguish close VAs even with large measurement errors (Fig. \ref{fig:S2A1SLAM}[c]). Figs. \ref{fig:S2A1SLAM}(d)-(f) show the detailed mapping performance of the three different simulation settings, respectively. Figs. \ref{fig:S2A1SLAM}(e) and \ref{fig:S2A1SLAM}(f) show that the mean absolute error (MAE) of VA2-3 decreases by $50\%$ with the help of IMU. We can observe that in scenario $2$, the birth and death phenomena of beams are considered, where VA2-1 disappears at $t=25$ and VA2-3 appears at $t=26$.
The MAE of VAs corresponding to PA $2$ is larger than that of PA $1$ because the UE location estimation deviates from the true value at $t=11$, resulting in poor initialization of VAs corresponding to PA $2$. 

\textbf{We can use acceleration measurements from IMU to characterize the turning of the UE, thus the IMU-Calib mechanism provides a more accurate track estimation than the non-IMU mechanism. The mapping performance is also enhanced. The results verify that IMU can improve the robustness of the proposed SLAM mechanism when radio measurements have poor quality.}

\subsubsection{Measurements}
We investigated the performance of the real measurements in scenario 2. Fig. \ref{fig:BA2}(a) shows that the measurements scattered from wall $1$ substantially deviate from the specular reflection because wall $1$ comprises multiple wooden doors, which have complex scattering characteristics. Given that the communication coverage in corridors is limited, the measurement errors are large at grid points $1$ and $2$, which are indicated by red or yellow colors for the three orientations in Fig. \ref{fig:BA2}. Moreover, the measurement errors are large at points that are behind the PA. The measurement accuracy from wall $2$, wall $3$, and LOS path is acceptable. However, the complex characteristics of scenario $2$ pose challenges to the proposed SLAM mechanism.

Figs. \ref{fig:S2A2SLAM}(a) and \ref{fig:S2A2SLAM}(b) show the SLAM results for real angle measurements without and with simulated IMU measurements, respectively. The legends ``w IMU'', ``w/o IMU'', and ``only IMU'' denote the IMU-assisted BP SLAM algorithm, the BP SLAM algorithm without IMU assistance, and the algorithm with only IMU and without BP SLAM, respectively.  w IMU and w/o IMU converge at around steps 15 and 10, respectively. However, only IMU does not converge and shows a drift phenomenon. The track estimation assisted by IMU (converges around 0.5 m) is more accurate than that without IMU (converges around 5 m) as shown in Fig. \ref{fig:S2A2SLAM}(c). Thus, the RSP estimates become increasingly accurate with the IMU assistance. Fig. \ref{fig:S2A2SLAM}(d) shows that the mean absolute error (MAE) of VAs decreases by $30\%$ on average with the IMU assistance. However, the MAE of VA2-1 is still large because the VA does not exist for diffuse reflections on wall $1$. Moreover, the estimations of VA2-1 exhibit dispersive properties, and the corresponding scattering points do not form a line segment.

\textbf{This experiment reveals that nonsmooth surfaces with complicated scattering characteristics cannot be characterized by anchor-like VAs and can be distinguished and filtered through the cluttered characteristics of the estimated scattering points.}

Challenges to the proposed SLAM mechanism posed by the complex characteristics of scenario $2$ include the following: 
(1) the phenomena of beam birth and death, as the UE moves, VA2-3 appears and VA2-1 disappears; 
(2) wall $1$ exhibits diffuse reflections; 
(3) the distance between glass decorative walls $4$ and $1$ is relatively close. 
Thus, distinguishing the two walls is difficult for the UE, which leads to data association errors and an increase in estimation errors for real measurements compared with simulations.

\textbf{The results in this section demonstrate that the proposed SLAM mechanism is effective in complex scenarios. The proposed SLAM mechanism can maintain its localization (with an error of ${1-2}$ m) and mapping (with MAE of around $5$ m) capabilities with nonsmooth scatterers and beam birth and death.}

\section{Conclusion}

This study investigated the angle-based SLAM mechanism in hardware-constrained mmWave systems without changing the communication transceiver architecture and 5G NR frame structure. First, we implemented a $28$ GHz mmWave ISAC platform and realized the beam management process according to 3GPP 5G NR standards. We then proposed a successive cancellation-based angle extraction approach to effectively retrieve AOAs and AODs from the real beam management process. We extended the classic BP SLAM algorithm to work only on angle measurements with single PA, tailored to the beam management process of mmWave networks with zero sensing overhead. Finally, we verified the proposed SLAM mechanism in empty hall and complex corridor scenarios, including multiple UEs or PAs, diffuse and specular scattering, and beam birth and death phenomena. 
The extensive results demonstrate the following. 
\begin{itemize}
\item The proposed SLAM mechanism is feasible for the realistic 5G NR mmWave system. The proposed mechanism can achieve sub-meter level localization and mapping accuracy in most experimental scenarios and retain localization (with an error of $1$-$2$ m) and mapping (with MAE of around $5$ m) capabilities in complex scenarios with nonsmooth scatterers, beam birth and death, and without the assistance of IMU.
\item Multiuser collaboration can help improve localization and mapping performance by approximately $50 \%$ and $77 \%$, respectively. 
\item IMU can help improve the mapping performance of the extreme VAs by approximately $50\%$ and $30\%$ for simulation and experiment, respectively.
\end{itemize}
Our future work will consider extending to 3D scenarios, unknown initial positions and orientations of PA and UE, and interference from multiple PAs and UEs.

\begin{appendices}
	\section{ }\label{A}
For data association, we define the data association vectors $ \bm{\psi}=[\mathbf{a},\mathbf{b}] $.
First, let $\mathcal{K}_{t-1}$ represent a set of legacy feature indexes.
The $|\mathcal{K}_{t-1}|$-dimensional feature-oriented vector is  ${\mathbf{a}_t=\big[{a}_{t,1},\ldots,{a}_{t,|\mathcal{K}_{t-1}|}\big]}$.
According to \cite{BP1}, the element of $\mathbf{a}_t$ is given by
\begin{equation}\label{DA1}
{a}_{t,i}\!=\!\left\{
\begin{array}{ll}
\!\!\!j \in \{ 1,\ldots, |\mathcal{M}_{t}|\}, &\!\!\text{legacy feature $i$ generates} \\
&\!\!\text{measurement $j$ at time $t$,} \\
\!\!\!0, &\!\!\text{legacy feature $i$ does not}\\
&\!\!\text{generate any measurement},
\end{array} \right.
\end{equation}where ${i = 1,\ldots, |\mathcal{K}_{t-1}|}$.
Second, the $|\mathcal{M}_{t}|$-dimensional measurement-oriented vector is ${\mathbf{b}_t=\big[{b}_{t,1},\ldots,{b}_{t,|\mathcal{M}_{t}|}\big]}$, and we obtain
\begin{equation}\label{DA2}
{b}_{t,j}\!=\!\left\{
\begin{array}{ll}
\!\!\! i \!\in \!\{ 1,\ldots, |\mathcal{K}_{t-1}| \}, & \!\!\text{measurement $j$ is generated} \\
& \!\!\text{by legacy feature $i$ at time $t$,} \\
\!\!\! 0, &\!\!\text{measurement $j$ is not} \\
&\!\!\text{generated by legacy feature,}
\end{array} \right.
\end{equation}
where ${j = 1,\ldots, |\mathcal{M}_{t}|}$.
Vectors $\mathbf{a}_t$ and $\mathbf{b}_t$, which are equivalent because one can be determined from the other,
can ensure the scalability properties of the BP algorithm.
A constraint exists such that each measurement originates from a maximum of one feature or one false alarm, and one feature can generate at most one measurement each time.
The exclusion-enforcing
function used to ensure the constraint is defined as
\begin{equation}\label{indicator}
\Psi(\mathbf{a}_{t},\mathbf{b}_{t})= \prod\limits_{i=1}^{|\mathcal{M}_{t}|}\prod\limits_{j=1}^{|\mathcal{K}_{t-1}|}\Psi({a}_{t,i},{b}_{t,j}),
\end{equation}
where
\begin{equation}
\Psi({a}_{t,i},{b}_{t,j}) = \left\{
\begin{array}{ll}
0,  &  {a}_{t,i}=j,\  {b}_{t,j} \neq i\ \\
  &\text{or} \  {b}_{t,j}=i,\  {a}_{t,i} \neq j,\\
1, &  \text{otherwise}.
\end{array} \right.
\end{equation}

\section{ }\label{B}
In this appendix, we introduce the classical BP SLAM algorithm \cite{rs4} and demonstrate how it can be used with AOAs and AODs in place of TOAs.

\subsubsection{State Transition}
Part (a) in \eqref{jointP1} denotes where the state transition function of UE is explained in Section \ref{imu}.
For the state transition function of feature $f(\tilde{{\mathbf{v}}}_{t,l}|{\mathbf{v}}_{t-1})$,
if a feature does not exist at the previous time, then it cannot exist as a legacy feature at the current time. Therefore,
for ${\tilde{{r}}}_{t-1,l}=0$, we obtain
\begin{equation}\label{tt1}
\begin{array}{ll}
&f(\tilde{{\mathbf{p}}}_{t,l},{\tilde{{r}}}_{t,l}|\mathbf{{p}}_{t-1,l},0)=\left\{
\begin{array}{lcc}f_{\rm D}(\tilde{{\mathbf{p}}}_{t,l}), & & {\tilde{{r}}_{t,l}=0},\\
0, & & {\tilde{{r}}_{t,l}=1},
\end{array} \right.
\end{array}
\end{equation}
where $f_{\rm D}(\cdot)$ is an arbitrary ``dummy" PDF, which is comprehensively explained in \cite{BP1}.
If a feature exists at the previous time, then the probability of its continuous existence at the current time is determined by the survival probability.
Therefore, for ${\tilde{{r}}}_{t-1,l}=1$, we obtain
\begin{equation}\label{tt2}
\begin{array}{ll}
&f(\tilde{{\mathbf{p}}}_{t,l},{\tilde{{r}}}_{t,l}|\mathbf{{p}}_{t-1,l},1)\\
&
=\left\{
\begin{array}{lc}
\left(1-P_{\rm s}(\mathbf{{p}}_{t-1,l})\right)f_{\rm D}(\tilde{{\mathbf{p}}}_{t,l}),  &  {\tilde{{r}}_{t,l}=0},\\
{P_{\rm s}(\mathbf{{p}}_{t-1,l})} f(\tilde{{\mathbf{p}}}_{t,l}|\mathbf{{p}}_{t-1,l}),  &  {\tilde{{r}}_{t,l}=1},
\end{array} \right.
\end{array}
\end{equation}
where ${P_{\rm s}(\cdot)}\in (0,1]$ represents the survival probability of a feature.

\subsubsection{Measurement Evaluation}
Part (b) in \eqref{jointP1} denotes where
a measurement can originate from a legacy feature, a new feature, or a false alarm.
In the angle-based SLAM, we first estimate RSPs and then obtain the corresponding VAs.
Therefore, we introduce $\mathbf{p}_{{\rm rsp},t}$ in the likelihood function to extend the BP SLAM algorithm \cite{rs4} by replacing TOAs with AOAs and AODs.
We define the likelihood function, that is, the PDF of measurements conditioned on the UE, features, RSPs, and data association vectors, as follows:
\begin{multline}\label{mmp}
f(\mathbf{z}_{t}|\mathbf{u}_{t},{\mathbf{v}}_{t},\mathbf{p}_{{\rm rsp},t},\mathbf{a}_{t},\mathbf{b}_{t})
= \!\!\! \prod\limits_{i\in\mathcal{D}_{t}}f(\mathbf{z}_{t,{a}_{t,i}}|\mathbf{u}_{t},\tilde{{\mathbf{v}}}_{t})\\
\times \prod\limits_{ j\in\mathcal{N}_{t}}\!\!\!f(\mathbf{z}_{t,j}|\mathbf{u}_{t},\breve{{\mathbf{v}}}_{t},\mathbf{p}_{{\rm rsp},t})
 \prod\limits_{q\in\mathcal{F}_{t}}f_{\rm false}(\mathbf{z}_{t,q}),
\end{multline}
where $\mathcal{D}_{t} \triangleq\left\{i \in\{1, \ldots, |\mathcal{K}_{t-1}|\}: a_{t, i} \neq 0\right\}$.
The likelihood function is updated to
\begin{multline}\label{mmmp2}
f(\mathbf{z}_{t}|\mathbf{u}_{t},{\mathbf{v}}_{t},\mathbf{p}_{{\rm rsp},t},\mathbf{a}_{t},\mathbf{c}_{t}) \\
\propto   \prod\limits_{i\in\mathcal{D}_{t}} \!\! \dfrac{f(\mathbf{z}_{t,{a}_{t,i}}|\mathbf{u}_{t},\tilde{{\mathbf{v}}}_{t})}{f_{\rm false}(\mathbf{z}_{t,{a}_{t,i}})}  \prod\limits_{j\in\mathcal{N}_{t}}  \dfrac{f(\mathbf{z}_{t,j}|\mathbf{u}_{t},\breve{{\mathbf{v}}}_{t},\mathbf{p}_{{\rm rsp},t})}{f_{\rm false}(\mathbf{z}_{t,j})},
\end{multline}
where the number of false alarms and newly detected features follows a Poisson distribution with a mean of $\mu_{\rm false}$ and $\mu_{\rm new}$, respectively.
The distribution of each false alarm measurement is described by the PDF $f_{\rm false}(\cdot)$.

\subsubsection{Data Association}\label{DA}
Part (c) in \eqref{jointP1} denotes where the joint prior PDF of data association vectors,  number-of-measurements vector, new features, and RSPs conditioned on the legacy features and the UE is $f(\mathbf{a}_{t},\mathbf{b}_{t},\mathbf{c}_{t},\breve{{\mathbf{v}}}_{t},\mathbf{p}_{{\rm rsp},t}|\tilde{{\mathbf{v}}}_{t},\mathbf{u}_{t})$.
According to Bayes' theorem, we have
\begin{multline}\label{dad}
f(\mathbf{a}_{t},\mathbf{b}_{t},\mathbf{c}_{t},\breve{{\mathbf{v}}}_{t},\mathbf{p}_{{\rm rsp},t}|\tilde{{\mathbf{v}}}_{t},\mathbf{u}_{t}) \\
= p(\mathbf{a}_{t},\mathbf{b}_{t},\mathbf{c}_{t},\breve{\mathbf{r}}_{t}|\breve{{\mathbf{p}}}_{t},\mathbf{p}_{{\rm rsp},t},\tilde{{\mathbf{v}}}_{t},\mathbf{u}_{t})\\
\times f(\breve{{\mathbf{p}}}_{t}|\mathbf{p}_{{\rm rsp},t},\mathbf{u}_{t})f(\mathbf{p}_{{\rm rsp},t}|\mathbf{u}_{t}).
\end{multline}
The joint prior pmf of the association vectors can be factorized as
\begin{multline}\label{da1}
p(\mathbf{a}_{t},\mathbf{b}_{t},\mathbf{c}_{t},\breve{\mathbf{r}}_{t}|\breve{{\mathbf{p}}}_{t},\mathbf{p}_{{\rm rsp},t},\tilde{{\mathbf{v}}}_{t},\mathbf{u}_{t}) \\
\propto   \Psi(\mathbf{a}_{t},\mathbf{b}_{t})(\mu_{\rm new})^{|\mathcal{N}_{t}|}(\mu_{\rm false})^{-|\mathcal{N}_{t}|-|\mathcal{D}_{t}|}\\
\times  \!\!\prod\limits_{i\in\mathcal{D}_{t}}\!\!\!P_{\rm d}(\mathbf{u}_{t},\mathbf{{p}}_{t,{a}_{t,i}}) \!\!\!\!\prod\limits_{i'\in\bar{\mathcal{D}}_{t}}\!\!\!\!\left(1\!-\!P_{\rm d}(\mathbf{u}_{t},\mathbf{{p}}_{t,i'})\right)\!\!\!\!
\\
\times \prod\limits_{j\in\mathcal{N}_{t}}\!\! \!\!f_{\rm new}(\breve{{\mathbf{v}}}_{t,j}|\mathbf{u}_{t})\!\!\!\!\prod\limits_{j'\in\bar{\mathcal{N}}_{t}}\!\!\!\! f_D(\breve{{\mathbf{v}}}_{t,j'}),
\end{multline}
where $\bar{\mathcal{D}}_{t}= \mathcal{K}_{t-1} \backslash {\mathcal{D}}_{t}$, $\bar{\mathcal{N}}_{t}= \mathcal{M}_{t} \backslash {\mathcal{N}}_{t}$, ``$\backslash$" represents the complement operator, $P_{\rm d}(\cdot) \!\!\in\!\! (0,1]$ is
the probability that a feature is ``detected" in the sense that it generates a measurement, and $f_{\rm new}(\cdot)$ represents the PDF of the newly detected features.

\subsubsection{Data Fusion}
The joint posterior PDF of the UE, features, and data association vectors conditioned on measurements for all $T$ times is $f(\mathbf{u}_{1:T}, {\mathbf{v}}_{1:T},\mathbf{p}_{{\rm rsp},1:T},\mathbf{a}_{1:T},\mathbf{b}_{1:T}|\mathbf{z}_{1:T})$.
According to \eqref{jointP1}, the joint posterior PDF is the product of state transition, measurement evaluation \eqref{mmmp2}, and data association \eqref{dad}.
Given that the factorizations of \eqref{mmmp2} and \eqref{da1}
are in the perspective of legacy and new features, we rewrite \eqref{mmmp2} and \eqref{da1} concisely.
${\mathcal{D}}_{t}$ is the set of legacy feature indexes that generate measurements at time $t$.
For $i \in {\mathcal{D}}_{t}$,
we have $\tilde{r}_{t,i} = 1$ and ${a}_{t,i}\neq 0$.
On the contrary,
$\bar{\mathcal{D}}_{t}$ is the set of legacy feature indexes that do not generate measurements at time $t$. For $i \in \bar{\mathcal{D}}_{t}$,
we have $\tilde{r}_{t,i} = 1$ and ${a}_{t,i}= 0$.
Thus, we define a function $g(\mathbf{u}_{t},\tilde{\mathbf{v}}_{t,i},\mathbf{a}_{t,i};\mathbf{z}_{t,i})$. When $\tilde{r}_{t,i} = 1$, we have
\begin{multline}\label{g1}	g(\mathbf{u}_{t},\tilde{\mathbf{v}}_{t,i},\mathbf{a}_{t,i};\mathbf{z}_{t,{a}_{t,i}})\\
=
\left\{
\begin{array}{ll}
\dfrac{f(\mathbf{z}_{t,{a}_{t,i}}|\mathbf{u}_{t},\tilde{{\mathbf{v}}}_{t,i})P_{\rm d}(\mathbf{u}_{t},\mathbf{{p}}_{t,i})}{\mu_{\rm false}f_{\rm false}(\mathbf{z}_{t,{a}_{t,i}})}, &   {a}_{t,i}\neq 0,\\
1-P_{\rm d}(\mathbf{u}_{t},\mathbf{{p}}_{t,i}), & {a}_{t,i}= 0.
\end{array} \right.
\end{multline}
When $\tilde{r}_{t,i}\!\! =\!\! 0$, we have
$
g(\!\mathbf{u}_{t},\!\tilde{\mathbf{v}}_{t,i}\!,\!\mathbf{a}_{t,i};\mathbf{z}_{t,{a}_{t,i}}\!)=1.
$
Moreover, ${\mathcal{N}}_{t}$ denotes the set of measurement indexes generated by new features, which means that we have $\breve{r}_{t,j}\!\!=\!\!1$ and ${b}_{t,j}\!\!=\!\!0$ for $j \in {\mathcal{N}}_{t}$. By contrast,
$\bar{\mathcal{N}}_{t}$ denotes the set of measurements that are not generated by new features, and we have $\breve{r}_{t,j}\!\!=\!\!0$ for $j \in \bar{\mathcal{N}}_{t}$.
We define a function $h(\mathbf{u}_{t},\breve{\mathbf{v}}_{t,j},\mathbf{p}_{{\rm rsp},t,j},\mathbf{b}_{t,j};\mathbf{z}_{t,j})$.
When $\breve{r}_{t,j}=1$, we have
\begin{multline}\label{h1}
\!\!h(\mathbf{u}_{t},\breve{\mathbf{v}}_{t,j},\mathbf{p}_{{\rm rsp},t,j},\mathbf{b}_{t,j};\mathbf{z}_{t,j})\\
=\!\!\left\{ \!\!\!\!
\begin{array}{ll}
  0,  &   \!\!\!\!  {b}_{t,j}\neq 0,\\
  \dfrac{\mu_{\rm new}f_{\rm new}(\breve{\mathbf{{v}}}_{t,j}| \mathbf{u}_{t})f(\mathbf{z}_{t,j}|\mathbf{u}_{t},\breve{\mathbf{v}}_{t,j},\mathbf{p}_{{\rm rsp},t,j})}{\mu_{\rm false}f_{\rm false}(\mathbf{z}_{t,j})}, &    \!\!\!\! {b}_{t,j}= 0.
\end{array} \right.
\end{multline}
When $\breve{r}_{t,j}\!\!=\!0$, we have
$
h(\mathbf{u}_{t},\breve{\mathbf{v}}_{t,j},\mathbf{p}_{{\rm rsp},t,j},\mathbf{b}_{t,j};\mathbf{z}_{t,j}) = f_D(\breve{\mathbf{{v}}}_{t,j}).
$
The joint posterior PDF is given by \eqref{final},
where (a), (b), and (c) correspond to the state transition, measurement evaluation, and data association phases, respectively.

\newcounter{TempEqCnt}
\setcounter{TempEqCnt}{\value{equation}}
\setcounter{equation}{34}
\begin{figure*}[ht]
	\begin{multline}\label{final}
	f(\mathbf{u}_{1:T},{\mathbf{v}}_{1:T},\mathbf{p}_{{\rm rsp},1:T},\mathbf{a}_{1:T},\mathbf{b}_{1:T}|\mathbf{z}_{1:T})
	\propto   \underbrace{\prod\limits_{t=1}^{T}f(\mathbf{u}_{t}|\mathbf{u}_{t-1})f(\tilde{{\mathbf{v}}}_{t,l}|{\mathbf{v}}_{t-1,l})}_{(a)}\\
	\times \prod\limits_{t=1}^{T}\left(\underbrace{
		\Psi(\mathbf{a}_{t},\mathbf{b}_{t}\!)}_{(c)}  \underbrace{\!\!\prod\limits_{i=1}^{|\mathcal{K}_{t-1}|} \!\! g(\mathbf{u}_{t},\tilde{\mathbf{v}}_{t,i},\!\mathbf{a}_{t,i};\mathbf{z}_{t,i})\! \!\prod\limits_{j=1}^{|\mathcal{M}_{t}|}\!\!h(\mathbf{u}_{t},\breve{\mathbf{v}}_{t,j},\!\mathbf{b}_{t,j};\mathbf{z}_{t,j})f(\breve{{\mathbf{p}}}_{t,j}|\mathbf{p}_{{\rm rsp},t,j},\mathbf{u}_{t})f(\mathbf{p}_{{\rm rsp},t,j} |\mathbf{u}_{t})}_{(b)}\right).
	\end{multline}
	\hrulefill
\end{figure*}
\setcounter{equation}{\value{TempEqCnt}}

\end{appendices}

\bibliographystyle{IEEEtran}
\bibliography{bibsample}

\end{document}